%% file: main.tex
\documentclass[12pt]{article}

\usepackage{newtxtext,newtxmath}
\usepackage{graphicx}
\usepackage{amsmath}
\usepackage[utf8]{inputenc}
\usepackage{subfiles}
\usepackage{svg}
\usepackage{authblk}
\usepackage[margin=1in]{geometry}
\usepackage{float}
\usepackage{caption}

\renewenvironment{abstract}
	{\quotation}
	{\endquotation}

\date{}

\makeatletter
\renewcommand{\fnum@figure}{\textbf{Figure \thefigure}}
\renewcommand{\fnum@table}{\textbf{Table \thetable}}
\makeatother

\usepackage{url}

\DeclareUnicodeCharacter{2009}{\,}
\DeclareUnicodeCharacter{03C0}{$\pi$}

\newcommand{\mVec}[1]{\boldsymbol{#1}}
\newcommand{\e}{\mathrm{e}}
\newcommand{\im}{\mathrm{i}}

\def\scititle{
	Attosecond Coherent Electron Motion in a Photoionized Aromatic Molecule
}
\title{\bfseries \boldmath \scititle}

\input{authors.tex}

\begin{document} 

\maketitle

\begin{center}
  \textbf{Abstract} 
\end{center}
\begin{abstract}
In molecular systems, the ultrafast motion of electrons initiates the process of chemical change.
Tracking this electronic motion across molecules requires coupling attosecond time resolution to atomic-scale spatial sensitivity.
In this work, we employ a pair of attosecond x-ray pulses from an x-ray free-electron laser to follow electron motion resulting from the sudden removal of an electron from a prototypical aromatic system, para-aminophenol.
X-ray absorption enables tracking this motion with atomic-site specificity. 
Our measurements are compared with state-of-the-art computational modeling, reproducing the observed response across multiple timescales.
Sub-femtosecond dynamics are assigned to states undergoing non-radiative decay, while few-femtosecond oscillatory motion is associated with electronic wavepacket motion in stable cation states, that will eventually couple to nuclear motion.
Our work provides insight on the ultrafast charge motion preceding and initiating chemical transformations in moderately complex systems, and provides a powerful benchmark for computational models of ultrafast charge motion in matter.
\end{abstract}

\noindent

The motion of electrons plays a fundamental role for scientific and technological applications, ranging from the earliest stages of chemical reactions~\cite{calegari_open_2023} to solar energy harvesting~\cite{ostroverkhova_2016}, light-driven catalysis~\cite{Stockman_2018} and next generation microelectronics~\cite{schultze_2014, freudenstein_2022}.
The timescale for the coherent motion of electrons is set by the energy splitting of electronic states. 
For valence excited states of small molecules, 
this energy splitting is on the scale of an electron volt~(eV), corresponding to electron motion on few-femtosecond~(fs) to attosecond~(as) timescales.
This has driven the technological development of ultrashort light pulses to study the dynamics of electrons in matter~\cite{nobel}.

\begin{figure}
\centering
\includegraphics[width= \textwidth]{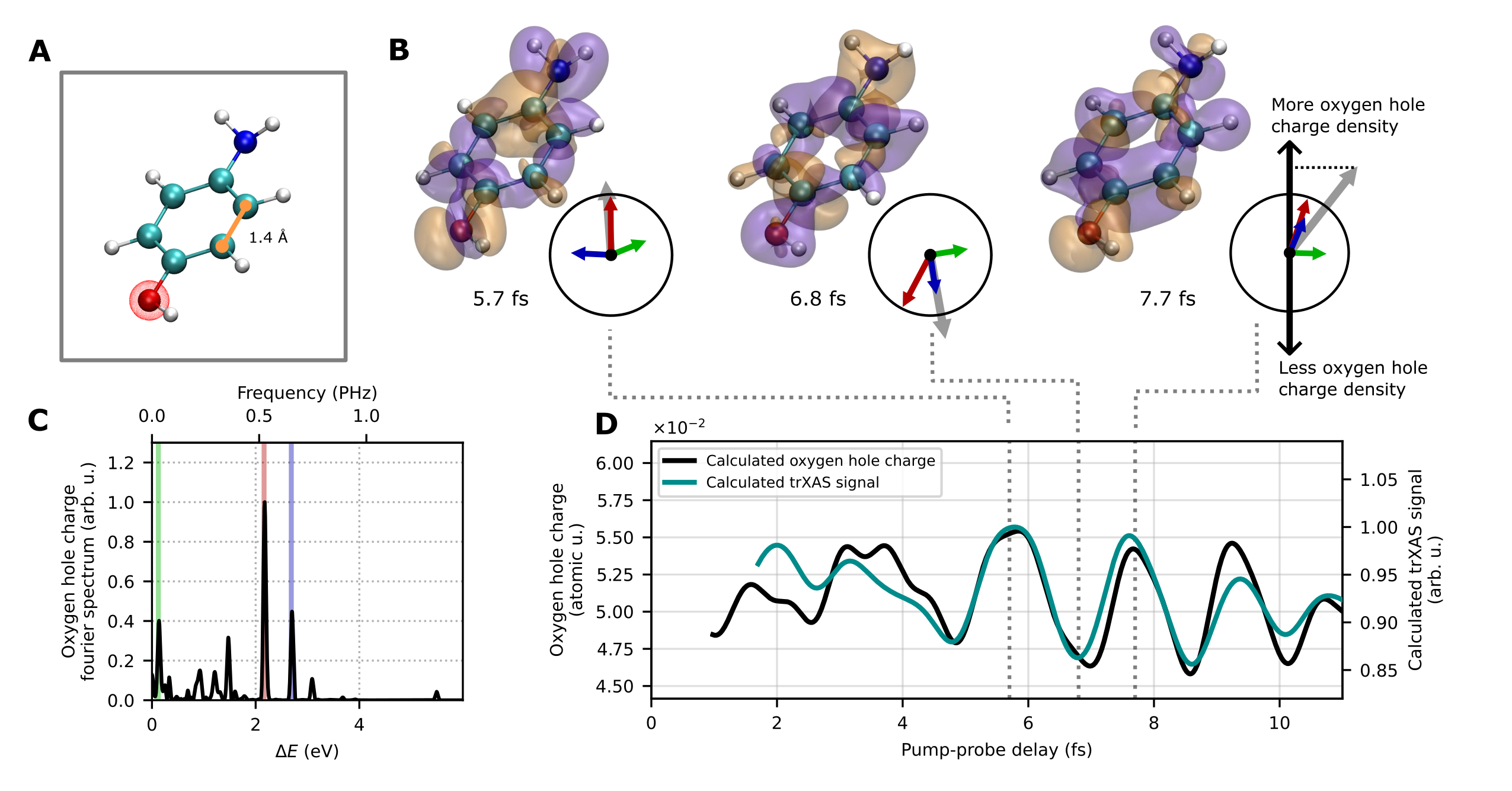}
\caption{\label{fig:fig1} \textbf{Simulation of ultrafast charge motion in para-aminophenol probed by trXAS at the oxygen $K$-edge.} (\textbf{A}) Molecular structure of para-aminophenol. Opaque spheres represent atomic positions: carbon (light blue), nitrogen (dark blue), oxygen (red), hydrogen (white). The translucent red sphere indicates the region of integration used to define the hole charge in panel \textbf{D}. 
 (\textbf{B}) Snapshots of the time-dependent part of the hole density~($\pm1 \times 10^{-4}$ isosurface), calculated using the RAS-SXD method (see main text for details). 
Purple represents a decrease in hole density (with respect to the static part of the hole density) and orange represents an increase.
The red, green and blue vectors show a representation in the complex plane of the three dominant frequency components involved in the motion (shown in panel \textbf{C}). The translucent black arrow is the vector sum of the three frequency components.
The real part of the black arrow~(projection onto the vertical axis) maps to the oxygen hole charge density.
(\textbf{C}) Fourier transform of the oxygen hole charge density shown in panel D. The red, green and blue vectors are the dominant frequency components, or cationic energy differences, involved in the motion. 
(\textbf{D}) Hole density in the vicinity of the oxygen atom~(hole density integrated over the red translucent sphere in panel \textbf{A}) compared to the simulated oxygen $K$-edge XAS signal in the probe photon energy range from 522--530 eV, calculated \emph{via} the RAS-SXD method (see main text). The curves do not begin at a pump-probe delay of zero, since the model does not access the dynamics in the region of overlap between the pump and probe pulses (see supplementary materials).}
\end{figure}

Coherently populating a superposition of eigenstates can initiate ultrafast motion in a quantum system. 
For example, impulsive excitation by an ultrashort light pulse can cause the charge distribution of a molecule to undergo rapid non-equilibrium motion~\cite{leone_what_2014}. 
For several decades, this motion has been approximated within the paradigm of `charge migration', in which motion is driven solely by a coherent excitation of ionized electronic states~\cite{cederbaum_ultrafast_1999}.
An example of this type of motion is calculated and shown in figure~\ref{fig:fig1}, which shows the response of the para-aminophenol molecule following impulsive valence ionization by an attosecond x-ray pulse.
Ionization by this pulse, which has a large bandwidth and can address a broad manifold of inner- and outer-valence orbitals, populates a superposition of delocalized electronic states in the cation. 
The coherent evolution of this superposition causes the hole density to migrate across the molecule (Fig.~1\textbf{B}), resulting in an oscillation of the hole density at the oxygen site (Fig.~1\textbf{D}).
In this calculation, the atomic nuclei are modeled as fixed-in-space.

The charge migration paradigm has motivated experiments on extreme timescales to understand the first moments of photochemistry~\cite{krausz_attosecond_2009}.
However, many aspects of attosecond electron motion remain unexplored~\cite{lepine_attosecond_2014,calegari_open_2023,sun_nuclear_2017,nisoli_attosecond_2017}.
To address this scientific challenge, we use a pair of attosecond x-ray pulses from a free-electron laser to resolve ultrafast electron motion with attosecond temporal resolution and \r{A}ngstrom-scale spatial sensitivity.
Figure~\ref{fig:fig1}\textbf{D} shows a comparison between the time-resolved x-ray absorption signal~(trXAS) below the oxygen $K$-edge~(523--529 eV, green curve) and the oscillating portion of the hole charge localized at the oxygen atom (black curves) following ionization by a sub-femtosecond ($700$~as full-width at half maximum), $\sim260$~eV x-ray pulse.
The two curves follow each other closely, demonstrating the power of x-ray observables to probe the local electronic density on sub-femtosecond timescales. 

Several pioneering measurements based on high harmonic generation~\cite{ferray_multiple-harmonic_1988} have provided insight into electron dynamics on the attosecond timescale.
A pair of ultraviolet~(UV) attosecond pulse trains has been used to probe electron dynamics in nitrogen molecules~\cite{okino_direct_2015} and a pair of single-femtosecond UV pulses was used to observe electron motion in xenon atoms~\cite{tzallas_extreme-ultraviolet_2011}.
Recently, a two-color UV attosecond pulse pair was employed to time-resolve the creation of the argon cation~\cite{kretschmar_compact_2024}.
Further measurements of ultrafast charge motion, which have extended to more complex molecular systems, have made use of strong-field interactions to initiate~\cite{goulielmakis_real-time_2010,kraus_measurement_2015,matselyukh_decoherence_2022} or interrogate~\cite{calegari_ultrafast_2014,kraus_measurement_2015,lara-astiaso_attosecond_2018,perfetto_ultrafast_2020,mansson_real-time_2021,mansson_ultrafast_2022} few-femtosecond scale charge motion in molecules.
The effect of a strong laser field on the molecular system is challenging to interpret because the interaction with the field is highly non-perturbative, and many observables rely on long-term outcomes such as molecular fragmentation, which introduces additional complexity to numerical modeling.
Comparatively, x-ray observables are more easily modeled and interpreted~\cite{chen_first-principles_2020,dutoi_time-resolved_2014,segatta_exploring_2019,neville_ultrafast_2018,cooper_analysis_2014}, which has been demonstrated in recent experiments using degenerate~\cite{schwickert_electronic_2022,schwickert_charge-induced_2022} or near-degenerate~\cite{barillot_correlation-driven_2021} few-femtosecond x-ray pulse pairs to probe electronic motion.
Importantly, x-rays also address core-to-valence transitions and therefore probe the valence electron density in the vicinity of the atomically-localized core-level electrons.

In this work, we use attosecond x-ray pulse pairs to access the sub-femtosecond electron dynamics following photoionization in the aromatic molecule para-aminophenol.
We probe the time-dependent electron density at the oxygen site \textit{via} trXAS at the oxygen $K$-edge, employing the yield of resonant Auger-Meitner electrons to measure the absorption intensity~\cite{cooper_analysis_2014}.
We compare our measurement with state-of-the-art computational modeling of the interaction between the molecule and both the pump and probe x-ray pulses.
Our study reveals the nature of the molecular response to impulsive ionization, which involves multiple timescales.
We observe sub-femtosecond population dynamics due to non-radiative, super-Coster-Kronig decay, and few-femtosecond oscillatory electron motion associated with stable cation states.
We attribute the overall decay of this oscillatory signal to the effect of nuclear motion on the few-femtosecond timescale, providing insight into the earliest stages of photochemistry in the studied molecule.

\section*{Experiment}
Two-color ($\omega/2\omega$) attosecond x-ray pulse pairs were generated at the Linac Coherent Light Source~(LCLS) XFEL using a harmonic configuration of the split-undulator method~\cite{guo_experimental_2024}, which is illustrated in Figure \ref{fig:figure2}\textbf{A}.
The delay between the pump~($\omega$) and probe~($2\omega$) pulses was controlled by changing the number of undulators used to produce the probe pulse, and adjusting the magnetic chicane that sits between the two undulator sections producing the pump and probe pulses~\cite{guo_experimental_2024}~(see supplementary materials).
The x-ray absorption signal was measured by recording the photon-energy dependent electron emission from the molecule using a magnetic bottle electron spectrometer~\cite{borne_design_2023}.
The photon-energy-resolved electron emission spectrum for the ground state molecule is shown in Fig.~\ref{fig:figure2}\textbf{C}.
The photoemission spectrum is dominated by several strong features with $\sim 470-510$~eV kinetic energy.
When the incoming photon energy is below the oxygen $K$-edge at 539~eV~\cite{zhaunerchyk_disentangling_2015}, these electrons are associated with resonant Auger-Meitner decay following $1s\rightarrow$valence absorption at $\sim532$~eV.
Above the oxygen $K$-edge, these electrons are produced by non-resonant ($KLL$) Auger-Meitner decay following $K$-shell ionization of the ground state molecule.
A much weaker signal from direct valence ionization can be seen as a faint, diagonal streak with $\sim 480-500$~eV kinetic energy at $510$~eV photon energy and increasing kinetic energy with increasing photon energy.
In our measurements, the total electron yield is directly proportional to the x-ray absorption.
To enhance the spectral resolution of our x-ray absorption measurement, we performed a shot-to-shot characterization of the incident x-ray spectrum coincident with the Auger-Meitner electron yield, using a variable line spacing~(VLS) grating x-ray spectrometer~\cite{obaid_lcls_2018} downstream of the interaction region, as shown in Fig.~\ref{fig:figure2}\textbf{A}.

The pump pulse was tuned to ${\sim}260$~eV, below the carbon $K$-edge, in order to drive ionization from the valence manifold.
We then probed the ultrafast electron motion by measuring the x-ray absorption spectrum in the vicinity of the oxygen $K$-edge, by scanning the central photon energy of the probe pulse from $500-530$~eV for each pump-probe delay. 
A simplified schematic of the experimental protocol is illustrated in Fig. 2\textbf{D}.
The electron hole created by the pump pulse produces a new feature in the oxygen $K$-shell absorption spectrum.
The strength of this absorption feature is predicted to depend on the localization of the hole at the oxygen atom, as demonstrated in Fig.~\ref{fig:fig1}\textbf{D}.
The core-excited ion produced by the core-to-valence absorption decays \textit{via} Auger-Meitner emission, emitting a broad spectrum of electrons with kinetic energies between $470-510$~eV.
The sample absorption is given by the yield of these Auger-Meitner electrons.
\section*{Experimental Results} 
\begin{figure}
\centering
\includegraphics{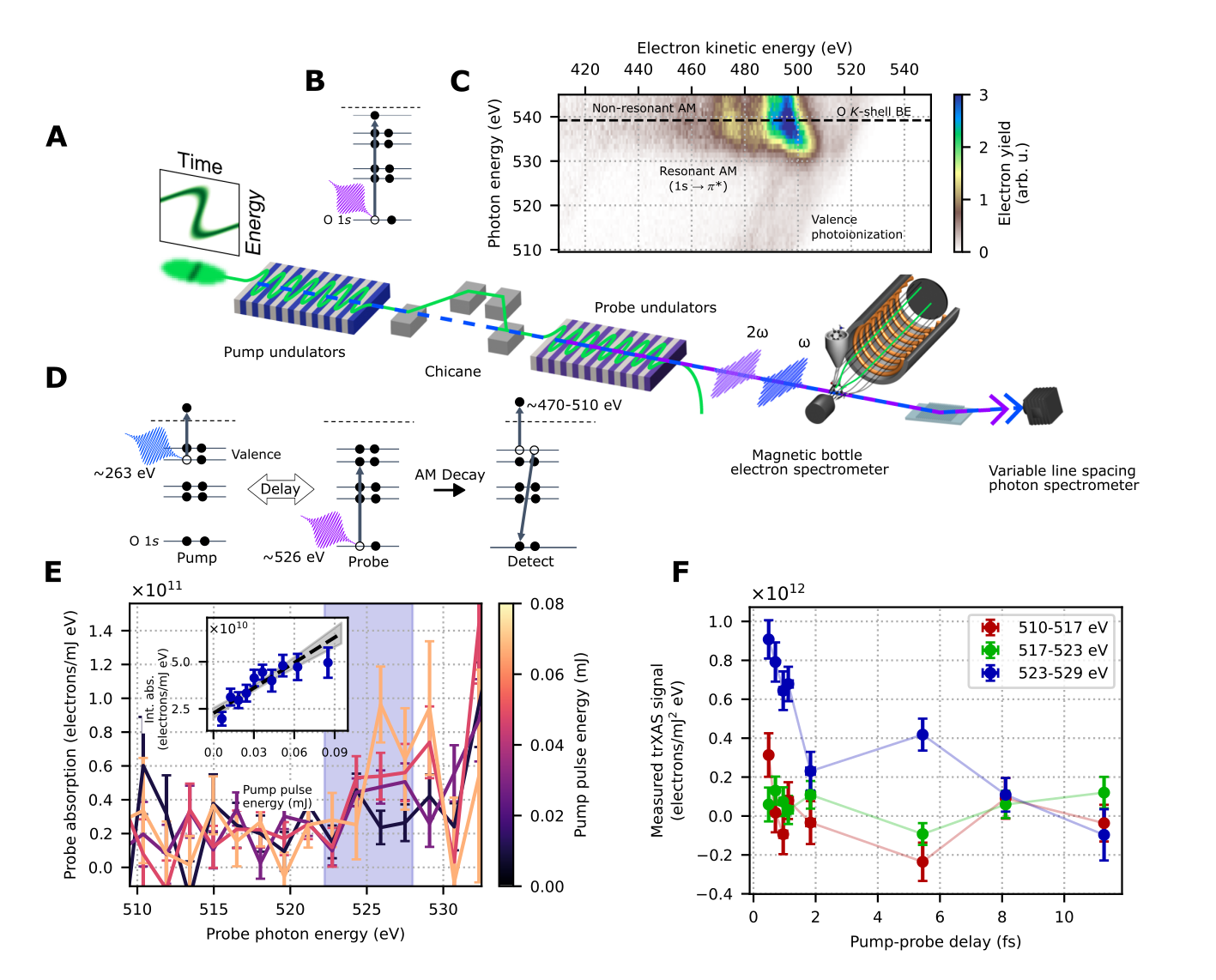}
\caption{\label{fig:figure2} \textbf{Measurement of ultrafast charge motion in para-aminophenol using attosecond x-ray pulse pairs.} (\textbf{A}) Experimental schematic showing the XFEL beamline, magnetic bottle electron spectrometer and x-ray spectrometer.
(\textbf{B}) Energy level diagram illustrating resonant x-ray absorption below the oxygen $K$-edge in the ground state molecule.
(\textbf{C}) Measured photon-energy resolved electron kinetic energy spectrum for the ground state molecule. The trXAS cross-section is calculated using the integrated electron yield in a kinetic energy range of 420--550~eV.
(\textbf{D}) Energy level diagram illustrating the pump, probe, and detection scheme used in this measurement. 
(\textbf{E}) Probe absorption as a function of probe photon energy (x-axis) and pump pulse energy (see colorbar), for a pump-probe delay of $5.4$~fs.
The inset shows the pump pulse energy dependence of the integrated yield in the blue shaded region.
(\textbf{F}) trXAS signal as a function of pump-probe delay for three different probe photon energy ranges.
}  
\end{figure}

Attosecond x-ray free-electron lasers rely on the process of self-amplified spontaneous emission~(SASE), where the x-ray pulse builds up from stochastic noise in the electron bunch~\cite{bonifacio_spectrum_1994}.
We exploit the resultant spectral fluctuations to achieve sub-bandwidth spectral resolution in the trXAS measurement using spectral domain ghost imaging~\cite{driver_attosecond_2020}.
Our analysis procedure is described in detail in the supplementary materials.
In order to extract the two-photon pump-probe signal, the data were binned according to the total pulse energy of the pump pulse. 
The x-ray absorption spectra for different pump energies are shown in Fig.~\ref{fig:figure2}\textbf{E} for a pump-probe delay of $5.4$~fs. 
A transient feature with a linear dependence on pump pulse energy appears in the probe absorption spectrum between $523$~to~$529$~eV, shaded in blue in Fig.~2\textbf{E}.
We extracted the pump-probe trXAS signal by regressing the total absorption in this region of interest on pump pulse energy, as shown in the inset of Fig.~\ref{fig:figure2}\textbf{E}. 
Negligible absorption was measured in the ground state molecule between $505$ and $530$~eV (this is shown as the gray curve in Fig.~\ref{fig:sim_res}\textbf{B}).

The delay-dependent pump-probe signal in the transient spectral feature, shown as the blue trace in figure~\ref{fig:figure2}\textbf{F}, decays rapidly within the first 2~fs and exhibits a revival around $5$~fs.
By approximately $10$~fs, the pump-probe signal has decayed to zero. 
We also analyzed the delay-dependence of the x-ray absorption below 523~eV~(green and red traces).
Between 517--523~eV~(green) we observed no delay-dependence.
Between $510$--$517$~eV~(red), we observed weak evidence of a sub-femtosecond decay, but the signal is very close to the noise floor of the measurement.

\section*{Discussion}
To understand the time-dependence of the pump-probe absorption signal, we compare our experimental results with \textit{ab initio} calculations of both the valence-induced electron dynamics and their associated x-ray absorption signatures.
We employed two state-of-the-art numerical approaches to evaluate the experimental observables following the interaction of the molecule with the pump and probe pulses, and
directly compare our experimental results with theoretical predictions.
We label the two numerical methods as ``ADC'' (algebraic diagrammatic construction) and ``RAS-SXD'' (restricted active space static-exchange B-spline density functional theory).
Both methods solve the time-dependent Schrodinger equation for the system interacting with the pump pulse.
Following the pump interaction, the photoelectron degree of freedom is traced out.
We then solve the von Neumann equation to calculate the evolution of the system until the end of the probe pulse.
The two methods are primarily differentiated by their description of the bound electronic states of the molecule, and the interaction terms used to model the effect of the pump pulse.
The ADC approach employs the time-dependent B-spline Restricted Correlation Space-Algebraic Diagrammatic Construction (RCS-ADC) method \cite{ruberti_restricted_2019,ruberti2019onset,ruberti_quantum_2022}. 
The time-dependent many-electron wavefunction is expanded on the basis of ground and excited RCS-ADC(2x) states, where the ``(2x)'' indicates the inclusion of the configuration space of two-hole, two-particle states.
This expansion incorporates terms that describe interchannel coupling between different ionic states as the photoelectron departs the cation.
Within the RAS-SXD approach we use multi-configurational restricted active space self-consistent field (RASSCF) wavefunctions to describe the bound molecular states, with dynamic correlation included perturbatively (RASPT2)~\cite{LiManni_JCTC_2023}.
The pump pulse populates single-channel continua, with the continuum orbitals described by static-exchange B-spline density functional theory (SXD)~\cite{Decleva_M_2022,Toffoli_CP_2002,Ponzi_JCP_2016}.
Both calculations are performed at the equilibrium geometry of neutral para-aminophenol.
The numerical methods are described in detail in the supplementary materials.

The calculated x-ray photoelectron binding energy spectrum produced by the pump pulse is compared to our experimental measurement in Fig.~\ref{fig:sim_res}\textbf{A}, where good agreement is found.
Fig.~\ref{fig:sim_res}\textbf{B} shows a comparison of measured and calculated trXAS spectra at short pump-probe delays~($<2$~fs).
Both methods predict that the absorption feature opened by the pump pulse contains comparable contributions from cationic states energetically below and above the double ionization potential~(DIP) of the molecule (21~eV), as shown in Fig.~\ref{fig:sim_res}\textbf{B}. 
There is significant energetic overlap, which is larger in the RAS-SXD model, between the absorption signal for these two sets of states. 
The states above the DIP mainly couple to higher-lying satellite states of the core-excited cation, and the energies of the strongest x-ray absorption transitions are found to be similar for states above and below the DIP.
States above the DIP are open to rapid autoionization \textit{via} super-Coster-Kronig decay~\cite{guillot_resonant_1977}, in which the vacancy is filled by an electron from the same shell, resulting in emission of another electron also from the same shell.
This autoionization process, which is illustrated in Fig.~\ref{fig:sim_res}\textbf{D}, typically occurs on the sub-femtosecond timescale and has not previously been resolved in the time domain.
We simulate the energy-dependent lifetimes for super-Coster-Kronig decay to be between $0.35-0.86$~fs with the Fano-ADC(2,2) method~\cite{kolorenc_fano-adc22_2020}, see supplementary materials for details.

The simulated delay-dependent XAS observable is shown in Fig.~\ref{fig:sim_res}\textbf{C}.
Both simulations predict few-femtosecond oscillations in the absorption signal, driven by the electronic coherence imparted by the pump pulse.
Strikingly, both the characteristic modulation frequency and the modulation depth are very different for the two methods.
We performed a detailed comparison of the ADC and RAS-SXD approaches to identify possible sources for this discrepancy.
The two methods employ different descriptions of the bound states and ionization continua, as described in the supplementary materials.
Both calculated pump photoelectron spectra agree well with our measurement.
This indicates that the photoionization cross sections, or the magnitude of the ionization amplitudes, do not strongly differ between the two calculations.
In contrast, phase differences between the ionization amplitudes will manifest as differences in the time-resolved signal.
These phase differences are sensitive to small differences in the electronic wavefunctions and rely critically on the description of both the bound and continuum wavefunctions.
The predicted time-domain signal may also be sensitive to the inclusion of interchannel coupling between the photoelectron and the cation, in the ADC simulation of the pump ionization step~\cite{ruberti_quantum_2022}.
This discrepancy between the time-domain predictions of our two state-of-the-art numerical approaches models highlights one advantage of time-domain measurements.

In Fig. \ref{fig:sim_res}\textbf{E} we directly compare the simulated trXAS signals with our experimental measurement.
We scale the absolute magnitude of the simulated trXAS signals to achieve the best overall agreement between experiment and theory for each numerical method.
We also model jitter in the pump/probe delay by convolving the curves shown in Fig.~\ref{fig:sim_res}\textbf{C} with a gaussian kernel of $\sigma=150$~as~\cite{guo_experimental_2024}, which generates the dashed lines in Fig.~\ref{fig:sim_res}\textbf{E}.
The signal within the region of overlap between the two pulses (delays shorter than the gray vertical line in Fig. \ref{fig:sim_res}\textbf{E}) is challenging to simulate, and is excluded from the RAS-SXD model.
In the ADC calculation, this region is accounted for by artificially truncating the pump pulse, which may neglect important processes contributing to the overall trXAS signal.
We attribute the increase in the measured absorption at 5.4~fs to a transient revival of valence hole density in the vicinity of the oxygen atom driven by coherent electron motion.
The magnitude of this revival in our experimental measurement agrees well with the ADC calculation and is larger than the RAS-SXD prediction.

The experimental measurement shows a rapid decay of the trXAS signal within the first femtosecond.
We partially attribute this rapid initial decay to the super-Coster-Kronig effect, which results in almost complete depopulation of states above the DIP within the first femtosecond.
This process is incorporated in our calculations as a phenomenological decay of the populations of the states above the DIP with the aforementioned lifetimes of 0.35~fs -- 0.86~fs.
The additional absorption from dication states populated through autoionization is also included in the simulated trXAS signal.
For longer delays above 1.7~fs, our simulations indicate that the x-ray absorption in the dication may partially compensate the decrease in the cation absorption.
We expect this rapid initial decay to be a general feature of the impulsive response of a molecule to ionization by broad-bandwidth light.

An additional feature in our experimental result is a decay in the overall XAS signal after roughly 5~fs, which is not well represented by either numerical method in Fig.~\ref{fig:sim_res}\textbf{E}.
This is likely a result
of nuclear motion, for example proton loss, which is not accounted for in our simulations and may shift or broaden the absorption feature shown in Fig.~\ref{fig:sim_res}\textbf{B}. 
Nonadiabatic electron-nuclear dynamics can also affect the absorption spectrum by modulating the excited state populations~\cite{wolf_probing_2017,vismarra_few-femtosecond_2024}.
We investigate this overall decay by incorporating an empirical decay term in the calculated XAS signal.
A simple exponential model can account for population decay, a shift or broadening of the absorption feature, and pure dephasing of a two-level system~\cite{vacher_electron_2017}.
We use the ADC calculation to find the overall decay constant that best matches our experimental measurement, which is 6.8~fs.
We find that incorporating this decay term for both simulated trXAS traces, shown in Fig.~\ref{fig:sim_res}\textbf{D} as the solid lines, improves the agreement between theory and experiment.
Our result highlights the need for complex modeling of effects such as nuclear motion and electron-nuclear coupling, to fully describe ultrafast charge motion on the sub- to few-femtosecond timescale.
    
\section*{Conclusion}
We have performed an attosecond x-ray pump/attosecond x-ray probe measurement of ultrafast charge motion in para-aminophenol and observed the impulsive response of a molecule to impulsive ionization.
Following interaction with a $260$~eV attosecond pump pulse, a transient feature was observed in the x-ray absorption spectrum roughly $\sim$20~eV below the oxygen $K$-edge of the neutral molecule.
By monitoring the delay-dependent yield of this absorption feature, we probed the time-evolving electron density in the molecule from the instant of photo-excitation with attosecond time resolution and atomic-site specificity.
Our work is a time-resolved experimental study of ionization in the impulsive limit, with sufficient photon energy and coherent bandwidth to create a broadband superposition of valence-ionized states including states above and below the DIP. 
We observe evidence for three distinct timescales: the decay of states above the DIP within the first femtosecond, the few-femtosecond coherent dynamics associated with coherently populated states of the cation, and the reduction of the signal over 5-10~fs, which we tentatively associate with the onset of nuclear motion. 
We expect this behavior to be general to small molecules, not only in the case of photoionization but also in response to fast ion or electron ionization of the valence shell, where the impulsive limit is reached.
Our work provides insight into the earliest stages of photoinduced chemistry and highlights the need to further explore the impact of coherent electronic phenomena on photochemical processes through the coupling of nuclear and electronic motion.
X-ray attosecond pump/probe spectroscopy provides the experimental framework to do this in a broad range of molecular systems, by achieving atomic specificity and attosecond time resolution.

\section*{Acknowledgments}
The authors gratefully acknowledge Jun Wang for expert support of the data analysis software.
\paragraph*{Funding:}
Use of the Linac Coherent Light Source (LCLS), SLAC National Accelerator Laboratory, is supported by the U.S. Department of Energy~(DOE), Office of Science, Office of Basic Energy Sciences~(BES) under contract no.~DE-AC02-76SF00515. 

The work performed by T.D., E.I., J.T.O., D.G., E.T., A.L.W., T.J.A.W., M.F.K., P.H.B., and J.P.C. were supported by the Chemical Sciences, Geosciences, and Biosciences Division (CSGB), BES, DOE. 
Z.G., D.C., J.D., S.L., K.A.L., P.L.F. N.S.S., R.R., Z.Z., and A.M. acknowledge support from the Accelerator and Detector Research Program of the Department of Energy, Basic Energy Sciences division. 
The work by G.A.M, D.T., Z.W., and L.F.D was supported by CSGB, BES, DOE under award no. DE-FG02-04ER15614 and DE-SC0012462. 
S.B. and N.B. acknowledge support from CSGB, BES, DOE under award no. DE-SC0012376. 
K.B. was supported by the Chemical Sciences, Geosciences, and Biosciences Division, Office of Basic Energy Sciences, Office of Science, US Department of Energy, Grant No. DE-FG02-86ER13491 and acknowledges additional support by the US Department of Energy, Office of Science, Office of Workforce Development for Teachers and Scientists, Office of Science Graduate Student Research (SCGSR) program. 
The SCGSR program is administered by the Oak Ridge Institute for Science and Education (ORISE) for the DOE. 
ORISE is managed by ORAU under contract number DE-SC0014664.
The work by G.G., A.Pa., and F.M. was funded through the European Research Council~(ERC) under the European Union’s Horizon 2020 research and innovation programme (grant agreement no. 951224, TOMATTO), the European COST Action CA18222 (AttoChem), supported by COST (European Cooperation in Science and Technology), the Comunidad de Madrid (project FULMATEN, Ref. Y2018 NMT-5028), the Spanish Ministerio de Ciencia e Innovación (projects PID2022-138288NB-C31 and PID2022-138288NB-C32, the ``Severo Ochoa'' Programme for Centres of Excellence in R\&D CEX2020-001039-S and the ``María de Maeztu'' Programme for Units of Excellence in R\&D CEX2018-000805-M). The calculations labeled RAS-SXD were performed at the Mare Nostrum Supercomputer of the Red Española de Supercomputación (BSC-RES) and the Centro de Computación Científica de la Universidad Autónoma de Madrid (CCC-UAM).
The work by M.R. and V.A. was supported by the EPSRC grant number EP/V009192/1.
A.Pi. acknowledges the Spanish Ministry of Science, Innovation and Universities \& the State Research Agency through grants refs. PID2021-126560NB-I00 and CNS2022-135803 (MCIU/AEI/FEDER, UE), and the ``Mar\'ia de Maeztu'' Programme for Units of Excellence in R\&D (CEX2023-001316-M).
K.B., E.W., D.R., and A.R. were supported by the U.S. DOE, Office of Science, Office of BES, CSGB under Award No. DE-FG02-86ER13491.''
The work by O.A. and J.P.M. was supported by UK EPSRC grants EP/X026094/1, EP/V026690/1 and EP/T006943/1
G.D and L.Y. were supported by the U.S. DOE, BES, CSGB under contract DE-AC02-06CH11357.
Work by O.G., D.S.S., M.S. and T.W. was supported by DOE BES CSGB under Award No. DE-AC02-05CH11231.

\paragraph*{Author contributions:}
T.D., P.H.B., M.F.K., J.P.M., P.W., A.M., J.P.C. conceived the research.
T.D., Z.G., J.T.O., O.A., D.C., J.D., D.G., K.A.L., S.L., G.A.M., D.T., N.B., C.B., K.B., X.C., L.F.D., G.D., P.L.F., A.K., X.L., M.F.L., R.O., R.R.R., D.R., A.R., D.S.S., N.S.S., E.T., K.U., E.W., A.L.W., T.W., T.J.A.W., L.Y., Z.Z., O.G., M.F.K., J.P.M., P.W., A.M., J.P.C. participated in collecting the data.
Z.G., D.C., J.D., S.L., P.L.F., R.R.R., N.S.S., Z.Z., A.M. generated the attosecond XFEL pulse pairs.
E.I., T.D., Z.G., J.T.O., O.A., S.B., D.C., J.D., D.G., K.A.L., S.L., P.K., G.A.M., D.T., Z.W., J.P.C. performed data analysis.
G.G., M.R., P.K., G.A.M., A.Pi., V.A., A.Pa., F.M. generated theoretical calculations.
All authors participated in interpreting the results and writing the manuscript.

\paragraph*{Competing interests:}
There are no competing interests to declare.
\paragraph*{Data and materials availability:}
Data will be made publicly accessible when the paper is published.

\subsection*{Supplementary materials}
Materials and Methods\\
Figs. S1 to S7\\
References \textit{(50-\arabic{enumiv})}\\

\begin{figure}
\centering
\includegraphics[width=0.65\textwidth]{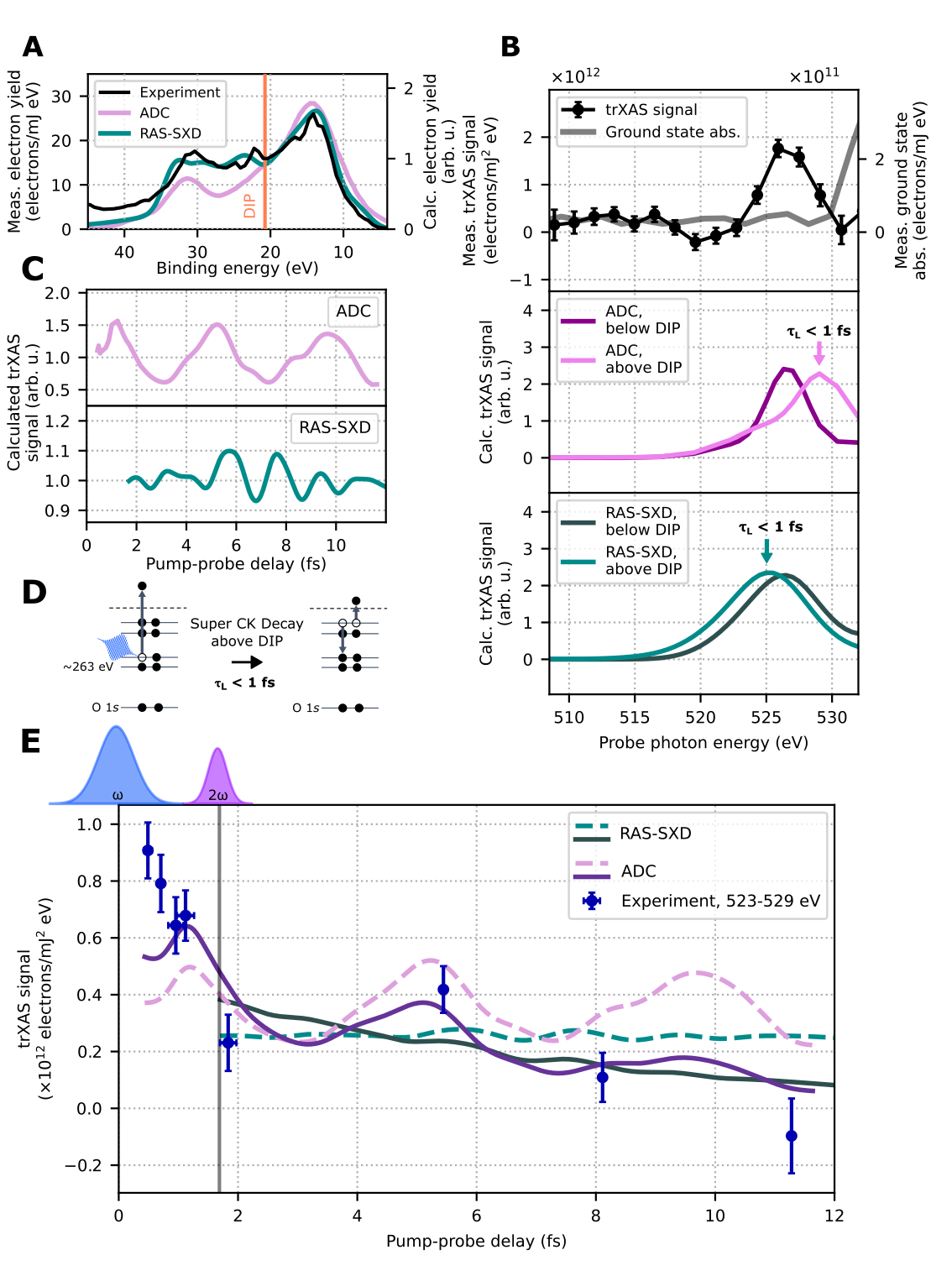}
\caption{\label{fig:sim_res} \textbf{Comparison between experimental measurement and simulations of experimental observable.}
(\textbf{A})~Simulated valence photoelectron spectrum produced by pump pulse compared to experimental measurement, where DIP is the double ionization potential~($\sim21$~eV).  
(\textbf{B}) Measured trXAS signal at $0.49$~fs pump-probe delay~(black line, upper panel) and ground-state absorption~(gray line, upper panel), calculated ADC trXAS signal at $1.05$~fs pump-probe delay (middle panel) and calculated RAS-SXD trXAS signal at $1.70$~fs pump-probe delay~(lower panel). 
The ground-state absorption overlaps with pump-probe signal (shown in gray) at higher photon energies ($>$530~eV). 
The calculated absorption signals are plotted separately for states above and below the DIP.}
\end{figure}

\makeatletter
\setlength{\@fptop}{0pt}
\makeatother

\begin{figure}
        \ContinuedFloat
        \centering
        \caption[]{
        \textbf{Comparison of Experimental Measurement and Simulated Observable, \textit{cont}.}\\[0.2em]
        As discussed in the text, the states above the DIP decay within less than 1~fs, as shown by the vertical arrows.
        (\textbf{C}) Calculated trXAS signal integrated across the photon energy range centered on 526 eV ($\pm 3$~eV for the ADC and $\pm 4$~eV for the RAS-SXD calculation). 
        (\textbf{D}) Schematic showing the non-radiative super-Coster-Kronig decay process for cationic states above the DIP.
        (\textbf{E}) Comparison between the calculated and measured trXAS signals. 
        Solid lines incorporate a phenomenological decay term to account for nuclear motion~(see text). 
        The intensity profiles of the pump~(blue) and probe~(purple) pulses are shaded in the top left for a pump-probe delay of 1.7~fs.}
\end{figure}

\clearpage

\renewcommand{\thefigure}{S\arabic{figure}}
\renewcommand{\thetable}{S\arabic{table}}
\renewcommand{\theequation}{S\arabic{equation}}
\renewcommand{\thepage}{S\arabic{page}}
\setcounter{figure}{0}
\setcounter{table}{0}
\setcounter{equation}{0}
\setcounter{page}{1} 

\newpage

\section*{Supplementary Material}

\setcounter{figure}{0}
\makeatletter 
\renewcommand{\thefigure}{S\@arabic\c@figure}
\makeatother

\setcounter{equation}{0}
\makeatletter 
\renewcommand{\theequation}{S\@arabic\c@equation}
\makeatother

\subsection*{Supplementary Materials for}

\subsection*{\scititle}

Taran~Driver~et~al. \break


\subsubsection*{This PDF file includes:}
Materials and Methods\\
Figs. S1 to S7\\
References \textit{(50-\arabic{enumiv})}\\ 
\newpage

\subsection*{Materials and Methods}

\section{Two-Pulse X-ray Generation}
Attosecond $\omega/2\omega$ pulse pairs with controllable delays down to the sub-femtosecond level were generated at the LCLS by lasing a high current spike in the split-undulator mode~\cite{guo_experimental_2024}.  with the harmonic lasing configuration. 
The high current spike in the electron bunch was produced by shaping the temporal profile of the photocathode laser~\cite{zhang_experimental_2020}. 
Single-spike attosecond soft x-ray pulses were generated by lasing this high current spike in the downstream undulator beamline. 
In the split-undulator mode, the same electron beam generates the pump and probe pulses in two different undulator sections, which are separated by a magnetic chicane~\cite{lutman_experimental_2013}.
Since the same electron bunch is used to generate the pump and probe pulses, the time delay between the pulses is independent of the electron beam arrival time jitter and therefore remains very stable~\cite{guo_experimental_2024}. 
The undulator parameters $K$ in the two undulator sections were tuned separately to produce attosecond $\hbar\omega{\sim}260~$eV and $2\hbar\omega{\sim}520~$eV pulses in the first and the second undulator sections, respectively. 
The time delay between the pump and probe pulses was controlled by tuning the magnetic chicane and the number of undulator modules used (i.e. set on resonance with the $2\omega$ probe pulse) in the second undulator section. 
In the harmonic lasing setup employed here, the strong microbunching produced in the high current spike during the generation of the pump~($\omega$) pulse was amplified in the generation of the probe~($2\omega$) pulse.
This ensured the power of the probe pulses was amplified to above GW-level in a short undulator distance, avoiding long slippage lengths and enabling sub-femtosecond pump-probe delays.

The time delays between pump/probe pulse pairs at a different photon energy setup ($370~$eV/$740~$eV) were measured in a separate experiment using the angular streaking technique~\cite{guo_experimental_2024}. 
We performed detailed time-dependent XFEL simulations to benchmark the time delays we measured in the angular streaking experiment and studied the dependence of the time delays on electron beam parameters, undulator beamline configurations, and XFEL photon energies. 
To calculate the time delays for our trXAS experiment, we employed the time-dependent XFEL simulations which we had previously benchmarked with  our angular streaking measurements.
We set the central energy of the electron bunch to 
$4.08~$GeV and adjusted the undulator matching condition to match the parameters used in the trXAS experiment, and simulated the average time delays for each beamline delay configuration that we used. 
The uncertainty in the simulated average time delays in the trXAS experiment was estimated by scaling the emittance of the electron bunch used in the numerical simulation. 
The maximal (minimal) emittance scaling ratio was determined by scaling the emittance until the simulated $\omega$ pulse energy reduced (increased) by a factor of 2.
Six values of emittance scaling ratio were taken between the maximal and minimal values, and an average time delay between $\omega/2\omega$ pulse pairs was calculated for each beamline configuration (i.e. pump/probe delay) for each emittance scaling ratio. 
The systematic uncertainty of the simulated average delay at a given beamline configuration, shown as the horizontal errorbar in Figs.~2\textbf{F} and~3\textbf{E} in the main text, was the standard deviation of these six simulated average delays.

\section{Experimental Setup}
The measurements were performed at the time-resolved molecular, and optical science~(TMO) experimental hutch of the LCLS~\cite{walter_time-resolved_2022}.
The x-ray pulses were focused using a pair of Kirkpatrick-Baez focusing mirrors~\cite{seaberg_x-ray_2022}.
The para-aminophenol sample was introduced using an in-vacuum oven heated to $\sim160~^{\circ}$C.
The focused x-ray beam intercepted the molecular sample in the interaction region of a two-meter magnetic bottle electron time-of-flight spectrometer~(MBES)~\cite{borne_design_2023}. 
For the trXAS measurements, a static $400$~V retardation voltage was applied to the flight tube of the MBES to increase the kinetic energy resolution for the Auger-Meitner electrons. 
The electrons were detected using a micro-channel plate detector coupled to a segmented anode.
The spectrum of the probe pulse was measured shot-to-shot by a variable line spacing~(VLS) grating spectrometer~\cite{obaid_lcls_2018}.
Since the gas-phase para-aminophenol sample was very dilute, the x-ray pulse experienced negligible attenuation through the sample and the spectral measurement made downstream of the interaction point was faithful to the incoming spectrum.

\section{Data Analysis}

We use spectral domain ghost imaging to retrieve the x-ray absorption cross section with sub-bandwidth resolution~\cite{driver_attosecond_2020}. 
In the present implementation, this technique is equivalent to multiple linear regression.
The natural variation of a self-amplified spontaneous emission~(SASE) XFEL, along with the variation induced by scanning the undulator $K$-value, creates shot-to-shot changes in the photon spectrum and we extract the correlation between the incident x-ray spectra and the response~(in this case, yield of resonant Auger-Meitner electrons) of the gas-phase sample. 
For a given pump pulse energy, the x-ray absorption is linearly related to the spectrum of the probe pulse. 
The corrected probe absorption signal (see section~\ref{sec:3w}) for a given pump pulse energy bin, $\Tilde{\Vec{b}}$, can therefore be written as the inner product of the single-shot photon spectrum, $\mathbf{A}$, with the unknown sample absorption, $\Vec{x}$: $\tilde{\Vec{b}} = \mathbf{A} \Vec{x}$.
We can estimate the sample absorption, $\Vec{x}$, by minimizing the squared difference between the estimated sample absorption and the measured sample absorption, given by 
\begin{equation}
    ||\textbf{A}_{i,\omega}\Vec{x}_{\omega} - \Tilde{\Vec{b}}_i||_2^2,
\end{equation}
where $\omega$ is the probe photon-energy index and $i$ is the shot index.
We removed shots with a central probe photon energy outside the range $509.5-532.5$~eV.
As a result, the mean spectrum of the probe pulse was similar across different delay points.
\begin{figure}
\centering
\resizebox{1\textwidth}{!}{\includegraphics{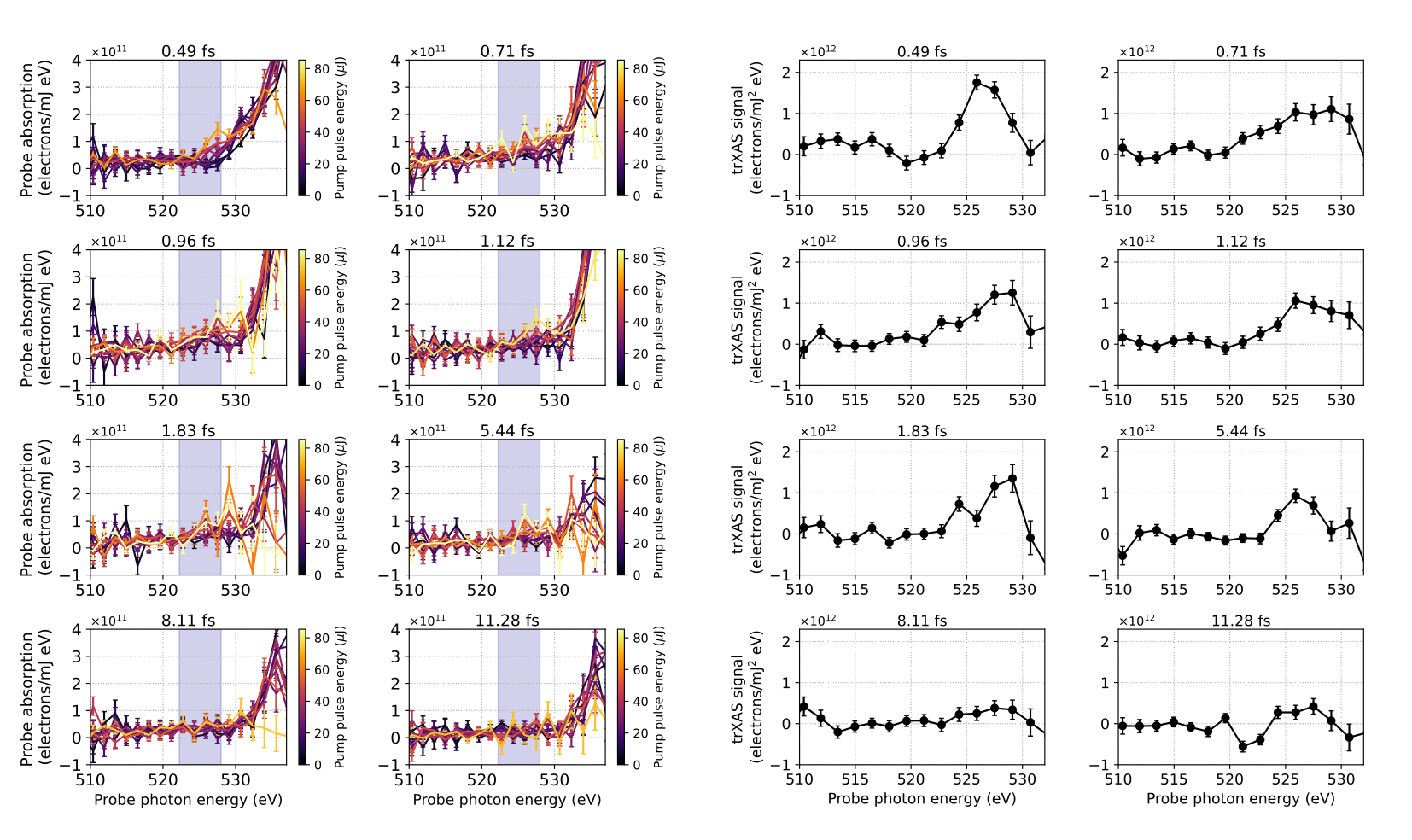}}
\caption{\label{fig:spook_results}Left side: Sample absorption reconstructed for each pump pulse energy bin, indicated by the colorbar on the side of each figure. 
The shaded blue region is the area of integration where the trXAS signal is observed.
Right side: Photon-energy-dependent trXAS signal regressed from the panels of the left side of the figure. 
The spectrum is resolved in photon energy, and has been smoothed by calculating a rolling average across two photon energy bins.
}
\end{figure}

\begin{figure}
\centering
\resizebox{1\textwidth}{!}{\includegraphics{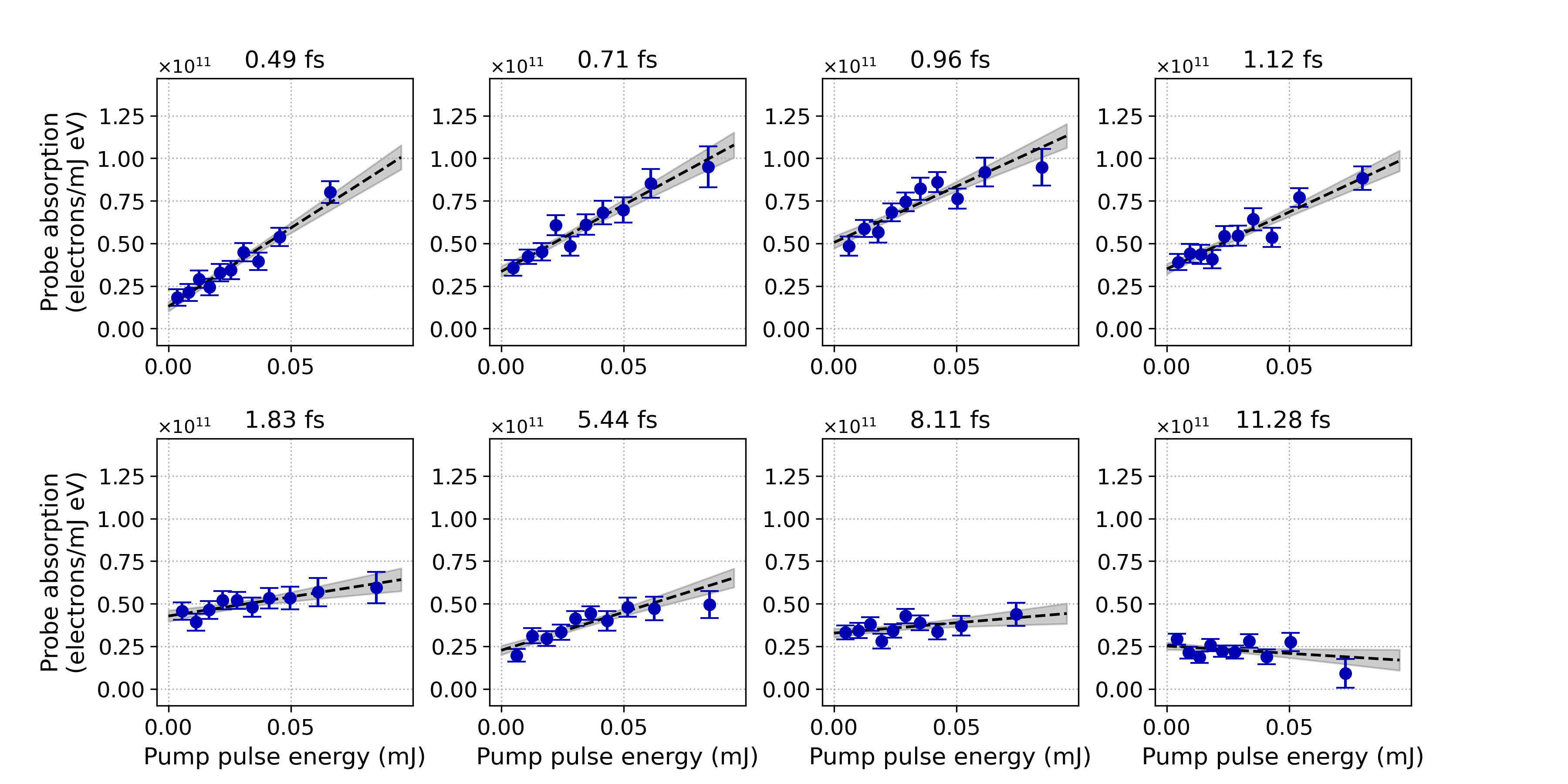}}
\caption{\label{fig:pump_reg}The integrated absorption yield is regressed on the pump pulse energy for each delay point and the pump/probe signal is extracted from the slope of the fit~(dotted line). The gray region is the one standard deviation confidence interval for the fit, given the error bars in the integrated absorption yield.}
\end{figure}

To evaluate the trXAS signal, which is bilinear in the intensity of the pump and probe pulses, the data is binned into 10 pump pulse energy bins such that each bin contains an equal number of shots and the sample absorption, $\Vec{x}$, is calculated for each pump pulse energy bin (left hand side of Fig.~\ref{fig:spook_results}). 
The absorption is then integrated across the resonance region from $523$~eV to $529$~eV, which is shown by the blue shaded region in Fig.~\ref{fig:spook_results}. 
Error bars for the integrated absorption yield are determined by re-sampling the data using bootstrap resampling (100 resamples). 
We then calculated the trXAS signal by calculating the slope of the absorption as a function of pump pulse energy, as shown in Fig.~\ref{fig:pump_reg}. 
Error bars in the trXAS signal are the standard error of the slope, given by the covariance matrix found when we regressed the integrated absorption on the pump pulse energy.
We enforced an upper limit on the probe pulse energy, by removing shots for which the probe pulse energy was higher than the value at which we observed evidence for saturation in our electron detection scheme.

\subsection{Contribution from Higher Harmonics of the XFEL}
\label{sec:3w}
\begin{figure}
\centering
\resizebox{\textwidth}{!}{\includegraphics{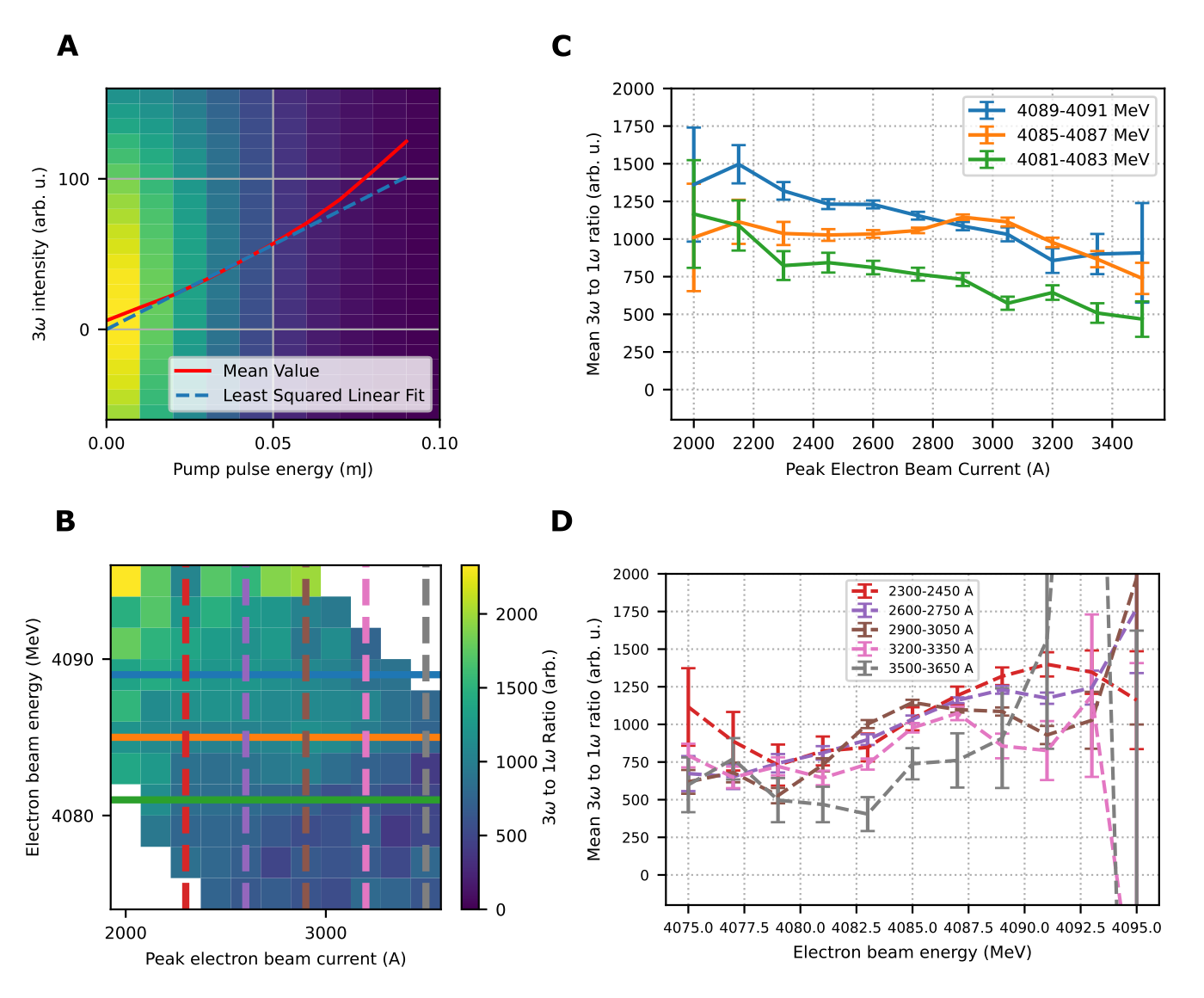}}
\caption{\label{fig:3w_char}\textbf{a)} Pulse energy of the third harmonic contribution measured using the variable line spacing spectrometer ($y$-axis), as a function of pump pulse energy ($x$-axis). Color map represents the number of shots measured at each value of third harmonic pulse energy and pump pulse energy.  
\textbf{b)} Ratio of pulse energies in third harmonic and pump pulse~(first harmonic), as a function of electron beam energy and peak electron beam current. \textbf{c)} Lineouts along the peak electron beam current axis while holding the electron beam energy fixed~(indicated by colored lines in panel \textbf{b}). \textbf{d)} Lineouts along the the electron beam energy axis while holding the peak electron beam current fixed~(indicated by colored lines in panel \textbf{b}).}
\end{figure}

Free-electron laser radiation can include a small contribution from higher order harmonics of the fundamental, most notably an on-axis contribution from the third harmonic at the sub-to-few percent level~\cite{ratner2011second}.
Our measurements of the x-ray spectrum and photoelectron spectrum confirmed that the first set of undulators also generated a small amount of on-axis radiation at the third harmonic of the pump photon energy~($3\omega{\sim}780$~eV).
We studied the behavior of the third harmonic pulse energy and found it to be approximately linear with the pump~($260$~eV) pulse energy, as shown in Fig.~\ref{fig:3w_char}A. 
Moreover, the pulse energy of the third harmonic also showed a weak correlation with energy of the electron beam, as shown in Fig.~\ref{fig:3w_char}D.

To quantify the contribution of the third harmonic to the measured Auger-Meitner electron signal, we recorded data with the the $K$-value of the probe undulators randomized, to suppress the lasing of the probe pulse.
We then measured the number of electrons produced by third harmonic in the same electron kinetic energy range analyzed in the experiment~($420-550$~eV), and using the same magnetic bottle electron spectrometer configuration.
The proportionality constant $\alpha(\gamma_{beam})$, between the pump pulse energy and the number of electrons produced by the third harmonic, was calculated for different electron beam energies~($\gamma_{beam}$).
For each shot in our pump/probe scan, we then performed a background subtraction using this measured proportionality constant, according to 
\begin{equation}
    \Tilde{\Vec{b}} = \Vec{b} - \alpha(\gamma_{beam}) E_{pump}
\end{equation}
where $\Vec{b}$ is the number of Auger-Meitner electrons recorded in a given shot, $\Tilde{\Vec{b}}$ is the value with the $3\omega$ background removed, and $E_{pump}$ is the pump pulse energy.
This correction was applied prior to the ghost imaging analysis.

\section{Numerical Methods}

\subsection{RAS-SXD simulation method}
The `RAS-SXD' abbreviation refers to the herein-employed combination of the restricted active space (RAS) quantum chemistry methods, to model multi-configurational bound states and O$(1s)$ absorption dipole moments of p-aminophenol, with the static-exchange B-spline density functional theory (SXD) method~\cite{Bachau_RPP_2001, Toffoli_CP_2002} to evaluate continuum orbitals and photoionization amplitudes.
We use the \textsc{openMolcas} software~\cite{LiManni_JCTC_2023} for the former, and the \textsc{Tiresia} code implemented by Decleva et al.~\cite{Decleva_M_2022, Toffoli_CPC_2024} for the latter.
Evaluating photoionization dipole moments in terms of SXD continuum orbitals and Dyson orbitals obtained from multi-reference bound-state quantum-chemistry methods has been introduced previously~\cite{Ponzi_JCP_2014, Ponzi_JCP_2016} and by now has been applied to a wide range of molecular systems
~\cite{Ponzi_JCP_2014, Ponzi_JCP_2016, Moitra_JCTC_2021, Delgado_FD_2021, Tenorio_M_2022}.
\subsubsection{Pump ionization and hole-charge dynamics}
We model pump and probe pulses as Gaussian fields centered at their half duration $T/2$,
\begin{align}
  \label{eq:sup:field}
  E(t) = E_0 \cos{\left[\omega \left(t-T/2\right)\right]} \e^{ -\frac{(t-T/2)^2}{2\sigma^2}}\,.
\end{align}

For the pump, we use a central frequency $\omega_1=$260~eV and an FWHM width of 0.70~fs, corresponding to a total duration of $T_1=1.96$~fs of $|E_1(t)|$, obtained from the 99\% percentile of the envelope.
The related FWHM-widths are 0.49~fs for $|E_1(t)|^2$, 5.2~eV for $|\tilde{E}_1(\omega)|$, and 3.7~eV for$|\tilde{E}_1(\omega)|^2$.
We model the dynamics of the created electron hole in an extension of the previously introduced single-configurational approach~\cite{Calegari_S_2014, Lara-Astiaso_FD_2016, Grell_PRR_2023}.
Briefly, the continuum character is traced out when the pump-pulse has decayed to zero at $T_1$, which yields a quantum mixture of bound cationic states.
The corresponding dynamic portion of the hole density (Fig.~1B) created by the pump in the residual cation is obtained as,
\begin{align}
  \label{eq:sup:dynHole}
  \rho^\text{dyn}_\text{hole}(\mVec{r},T_1+t) = -\sum_{i\neq j}\gamma^+_{ij}(T_1) \sum_{pq} \rho^{1e}_{ji;pq} \, \phi_p(\mVec{r}) \phi_q(\mVec{r})\e^{-i\mathcal{E}^+_{ij}t}\,.
\end{align}
The cationic density matrix $\gamma^+_{ij}(T_1)$ contains the pump-photoionization amplitudes for the cation states $i,j$ and takes into account the orientational average of the molecules with respect to the beam.
It is connected to the cation molecular orbitals $\phi_p(\mVec{r})$, $\phi_q(\mVec{r})$  by means of the spin-summed one-electron transition density matrix $\rho^{1e}_{ji;pq}$.
The oscillation frequencies are the cationic energy differences $\mathcal{E}^+_{ij}$.
The localized dynamic hole charge plotted in Fig.~1D has been obtained by integrating the dynamic hole-density in Eq.~\eqref{eq:sup:dynHole} over a sphere with $r=1.0$~a.u. centered on the oxygen atom of p-aminophenol.
\subsubsection{Transient pump-probe XAS signal}
\label{sec:sup:RAS:trXAS}
The O($1s$) transient XAS (trXAS) is evaluated by solving the von Neumann equation for the time evolution of the mixed cationic state during the presence of the probe pulse $E_2(t)$ to second-order of time-dependent perturbation theory under the assumption of non-overlapping pump and probe pulses,
\begin{align}
  \label{eq:sup:XAS}
  P^\text{XAS}_\text{O(1s)}(t,\omega_2) = \frac{2\pi}{3} \sum_{i,j}^{N_\text{val}} \sum_f^{N_\text{O(1s)} }  \sum_{\alpha=x,y,z}
                                     \gamma^+_{ij,\alpha}(T_1) \e^{-\im\mathcal{E}^+_{ij}\,t}
                                     \mu^+_{fi,\alpha} \mu^+_{jf,\alpha}
                                     \tilde{E}^*_2(\omega_2,\mathcal{E}^+_{fi}) \tilde{E}_2(\omega_2,\mathcal{E}^+_{fj})\,.
\end{align}
$\mu^+_{fi,\alpha}$ is the $\alpha$ component of the O(1s) absorption transition dipole moment between cation valence state $i$ and  O(1s)-hole state $f$.
$\mathcal{E}^+_{fi}$ denotes the respective energy difference absorbed from the probe field with central frequency $\omega_2$ and spectrum $\tilde{E}_2(\omega_2,\omega)$, defined analogously to the pump pulse, Eq.~\eqref{eq:sup:field}.
Since pump and probe pulses share the same polarization, the average over molecular orientations has been accounted for by averaging over the $x-x$, $y-y$, and $z-z$ combinations of pump and probe dipole components.
$t$ denotes the field-free time between pulses and is related to the pump-probe delay between the maxima of the pulses as $\tau = t + \frac{T_1+T_2}{2}$.
We have employed probe pulses with a spectral FWHM of 5.0~eV of $|\tilde{E}_2(\omega)|^2$, corresponding to the FWHM durations 0.37~fs for $|E_2(t)|^2$, 0.52~fs for $|E_2(t)|$ and a total duration of $T_2$=1.44~fs.
The minimum accessible delay is thus $\frac{T_1+T_2}{2}$=1.7~fs.
The trXAS has been obtained for central probe energies from 505~eV to 555~eV.
To obtain the time-dependent traces, the so-obtained trXAS has been integrated over the maximum between 522~eV and 530~eV.
\subsubsection{Autoionization of cation states}
\begin{figure}
  \centering
  \resizebox{1\textwidth}{!}{\includegraphics{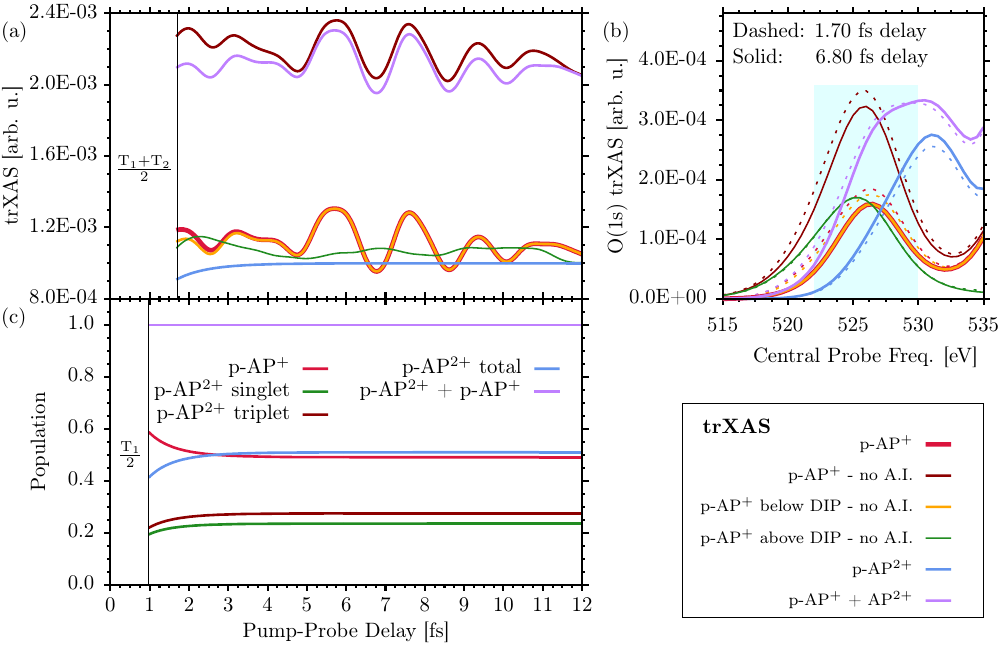}}
  \caption{
  \label{fig:RAS-SXD-trXAS}
  (a) trXAS traces obtained with the RAS-SXD method for the p-aminophenol cation (p-AP$^+$) and dication (p-AP$^{2+}$).
  The cation trXAS without autoionization (no A.I.) and its decomposition into contributions from states below and above the DIP are also indicated.
  (b) Corresponding photon-energy resolved trXAS at a delay of 6.8~fs (solid lines) compared to the minimum delay result at $\frac{T_1+T_2}{2}$=1.7~fs (dashed lines).
  The integration region of 522-530~eV, used to obtain the data in (a), is denoted with a light-blue rectangle.
  (c) Population dynamics of the p-aminophenol cation and dication species.
  }
\end{figure}

Within the RAS-SXD approach, we obtained double-ionization potentials (DIPs) of the lowest lying singlet, 20.74~eV, and triplet, 21.96~eV, dication states at the RASPT2 level.
For smaller molecules, valence autoionization lifetimes in the range of 1~fs to few fs have been reported~\cite{Feifel_PRL_2016}.
Herein, the autoionization lifetimes of cationic states above the DIP have been estimated with the Fano-ADC(2,2) method~\cite{kolorenc_fano-adc22_2020}, 
as
\begin{align}
  \label{eq:sup:decayFano}
  \frac{1}{\Gamma_i}= \tau_i =
    \left\{
      \begin{array}{rl}
        \text{0.345~fs}, & \text{DIP}\leq\mathcal{E}_{ig} < \text{23~eV}\\
        \text{0.425~fs}, & \text{23~eV}\leq\mathcal{E}_{ig} < \text{30~eV}\\
        \text{0.864~fs}, & \text{30~eV}\leq\mathcal{E}_{ig} < \text{34~eV}
      \end{array}
    \right.\,,
\end{align}
suggesting a sub-fs time-scale for the cation autoionization.
The RAS-SXD approach yields 2226 cation valence states, out of which 125 states lie below the DIP of 20.74 eV.
The autoionization of the states above the DIP is modeled as an phenomenological decay attached to the cation states during the pump interaction, free propagation, and probe interaction.
After the continuum part has been traced out, this introduces the following time-dependence to the density matrix elements, `autoionization' denoted by superscript `AI',
\begin{align}
  \rho^\text{+,AI}_{ij}(T_1+t) = \gamma^\text{+,AI}_{ij}(T_1) \e^{-\im\mathcal{E}^+_{ij} t}
  \e^{-\frac{\Gamma_i+\Gamma_j}{2} t}\,.
\end{align}
Including the 362 (166 singlet and  196 triplet) dication states below 35~eV to keep the computational costs under control, the cation-dication autoionization population dynamics is modeled by a master-equation driven by the Fano-ADC(2,2) decay rates, $\dot{\rho}^{2+}_{aa}(t) = \sum_i \Gamma_{i,a} \rho^\text{+,AI}_{ii}(t)$,
which has the solution,
\begin{align}
  \label{eq:sup:DicatPop}
  \rho^{2+}_{aa}(T_1+t) &= \rho^{2+}_{aa}(T_1) + \sum_i \frac{\Gamma_{i,a}}{\Gamma_i} \rho^\text{+,AI}_{ii}(T_1)\left(1-\e^{-\Gamma_i t}\right)\,.
\end{align}
Assuming equal probabilities of all (singlet and triplet) open dication energies, the partial decay rates are estimated in terms of the total decay rate multiplied with the ratio of the energy interval occupied by the dication state $a$, $\Delta\mathcal{E}^{2+}_a$, and the total range of open dication state energies,
\begin{align}
  \label{eq:sup:partRate}
  \Gamma_{i,a} = \Gamma_i\cdot\frac{\Delta\mathcal{E}^{2+}_a}
                              { \sum_{b;\mathcal{E}^{2+}_b<\mathcal{E}^+_i} \Delta\mathcal{E}^+_b }\,.
\end{align}
The dication populations created during the pump-pulse are obtained from the difference of the cation populations with and without autoionization,
\begin{align}
  \label{eq:sup:DicatPopInit}
  \rho^{2+}_{aa}(T_1) = \sum_i \left[\rho^+_{ii}(T_1) - \rho^\text{+,AI}_{ii}(T_1)\right] \frac{\Gamma_{i,a}}{\Gamma_i}\,,
\end{align}
ensuring the conservation of norm.
The trXAS with autoionization effects, exemplified for the cation, is obtained analogously to Eq.~\eqref{eq:sup:XAS} as,
\begin{align}
  \label{eq:sup:AI-XAS:cat}
    \begin{split}
      P^\text{AI,XAS}_\text{+,O(1s)}(t,\omega_2)=&
      \frac{2\pi}{3} \sum_{i,j}^{N^+_\text{val}} \sum_f^{N^+_\text{O(1s)} }
      \sum_{\alpha=x,y,z}
      \overbrace{\gamma^\text{+,AI}_{ij,\alpha}(T_1) \e^{-[\im\mathcal{E}^+_{ij} + \frac{1}{2}(\Gamma_i+\Gamma_j)]\cdot t}}
      ^{\rho^\text{+,AI}_{ij,\alpha} (T_1+t)}
      \mu^+_{fi,\alpha} \mu^+_{jf,\alpha}\\
      & \qquad \qquad \quad\times \mathcal{F}\left[ E_2(\omega_2,t)\e^{\Gamma_i t} \right]^*(\mathcal{E}^+_{fi})
        \mathcal{F}\left[ E_2(\omega_2,t)\e^{\Gamma_j t} \right](\mathcal{E}^+_{fj})
    \end{split}\,,
\end{align}
with $\mathcal{F}[\,\cdot\,]$ denoting the Fourier transformation.
The cation-diacation population dynamics, as well as the corresponding decomposition of the total trXAS signal into the cation and dication components is shown in Fig.~\ref{fig:RAS-SXD-trXAS}.
\subsubsection{Computational details}
\paragraph{RAS bound state calculations}
The equilibrium geometry has been optimized  with the \textsc{Gaussian16} program~\cite{Frisch2016} at the DFT level using the B3LYP functional and \textit{cc-pVTZ} basis set.
The neutral, cationic, and dicationic bound states have been computed with the restricted active space self-consistent field (RASSCF) method in \textsc{openMolcas}~\cite{LiManni_JCTC_2023} using the \textit{ANO-L TZP} basis set.
The active space used for all bound-state calculations comprises the O($1s$) orbital in RAS1, allowing for a single hole;
RAS2 contains all 21 valence orbitals at full flexibility;
RAS3 holds the lowest 5 virtual orbitals, allowing for a single electron.
All manifolds of states for all charge species were obtained in independent state-averaged RASSCF calculations over all configurations,
106 neutral, 2226/211 cation valence/O(1s), 15631/2226 dication singlet, and 22260/3276 dication triplet valence/O(1s).
To obtain the cation and dication O($1s$)-hole states, the `HEXS' keyword of the \textsc{openMolcas} RASSCF implementation has been used to remove configurations with a doubly occupied O($1s$) orbital from the CI expansion.
Further, all core-orbitals have been frozen to keep the  O($1s$) orbital in RAS1 during the optimization.
The cation valence manifold yields the tentatively assigned RAS3 orbital character $2\times \pi^*$, $\sigma(\text{Ryd}, 3s)$, and $2\times\sigma(\text{Ryd}, 3p)$.
The obtained active spaces for the other manifolds of states are similar, however, the amount of admixed Rydberg character in RAS3 decreases with increasing charge.
The RASSCF energies have been perturbatively corrected for dynamic correlation with the single-state RASPT2 method~\cite{Malmqvist_JCP_2008}, using an IPEA shift of 0.25 a.u. and an imaginary shift of 0.5 a.u.
The Dyson orbitals for the pump-ionization have been evaluated with the Spherical Continuum for Auger-Meitner decay and Photoionization (SCAMPI) code~\cite{Grell_JCP_2020, Grell_scampi_2024}.
The O($1s$) absorption dipole moments have been evaluated between the valence and O($1s$)-hole state manifolds of the cation and dication with the RASSI method~\cite{Malmqvist_CPL_2002}.
In that the customary rediagonalization of the Hamiltonian, constructed from the joint valence and O($1s$)-hole state manifolds, has not been carried out to avoid mixing both manifolds, which would considerably complicate the propagation of the dynamics.
We expect this to be a reasonable assumption for the present study since the residual overlap matrix elements between the valence and O($1s$)-hole states are comparatively small.
\paragraph{SXD continuum calculations}
Details of the static-exchange B-spline DFT method implemented in \textsc{Tiresia}~\cite{Decleva_M_2022, Toffoli_CPC_2024} can be found elsewhere~\cite{Bachau_RPP_2001, Toffoli_CP_2002}.
A reference ground-state DFT electron density of p-aminophenol was obtained with the LB94 functional in the basis of \textit{TZP} Slater-type orbitals with the ADF package~\cite{teVelde_JCC_2001}, which was then used to evaluate the LB94 static-exchange DFT Hamiltonian in the B-spline basis.
The basis has been build up from a one-center radial expansion of B-splines, spanning 22.5~a.u. with a 0.15~a.u. step, originating at the center of the aromatic ring with angular functions constituted by spherical harmonics of angular momenta up to $l_\text{max}=33$.
This expansion was supplemented by small multi-center B-spline expansions at the atomic positions with $l_\text{max}^\text{H}=1$ (hydrogen) and $l_\text{max}^\text{O,C,N}=2$ (other atoms).
The atomic expansions range 0.6 a.u. (carbon) and 0.7~a.u. (other atoms), respectively.
Photoionization dipole moments have been evaluated between all $(l^\text{max}+1)^2 = 1156$ degenerate continuum orbitals obtained at 51 kinetic energies spanning 5.0-10.0 a.u. in 0.1. a.u. steps and the 2226 pump-ionization Dyson orbitals projected to the B-spline basis~\cite{Ponzi_JCP_2014,Ponzi_JCP_2016}.

\subsection{\label{sec:lev1}ADC theoretical methods}

Theoretical modeling and simulation of the photoionization dynamics triggered by the x-ray laser pulses in the para-aminophenol molecule are performed for the method labeled `ADC' within the framework of the time-dependent B-spline restricted correlation space (RCS) – algebraic diagrammatic construction (ADC) \textit{ab initio} methods \cite{ruberti_restricted_2019,ruberti2019onset,Ruberti2021,Ruberti2023}. Both interactions of the neutral para-aminophenol molecule with the pump and of its cation with the probe x-ray attosecond pulses, respectively, are described from-first-principles at the equilibrium geometry of neutral para-aminophenol. 

\subsubsection{Simulation of x-ray attosecond pump ionization}

Within the B-spline RCS-ADC method, a multicenter B-spline basis set \cite{Toffoli_CP_2002,ruberti_restricted_2019} is used to describe single-electron states. The space of the unoccupied states resulting from a Hartree-Fock (HF) calculation is partitioned into two orthonormal subspaces: the restricted correlation space (RCS) ($\chi_{\alpha}$), designed to accurately describe the localized short-range correlation of the system, and its orthonormal complement, the ionization space (IS) ($\psi_{\mu}$), which describes the long-range part of the photoelectron wavefunction and complements the RCS space in order to represent the electronic wavefunction throughout the entire spatial region \cite{ruberti_restricted_2019}. Here and in what follows, the 
$\alpha, \beta, ....$ and $\mu, \nu, ....$ indices refer to the virtual space (unoccupied) RCS and IS orbitals, respectively, whereas $i, j, k, ....$, indicate the occupied HF molecular orbitals. In the simulation, we used a B-spline basis characterized by a radius of the radial box $R_{max} = 80$ a.u., a linear radial grid with step size $h = 0.45$ a.u., and a maximum value of orbital angular momentum $L_{max} = 12$. The RCS orbital space consists of the virtual orbitals from an HF calculation performed in the cc-pVDZ GTO basis set.

Within the time-dependent (TD) B-spline RCS-ADC(n) approach to molecular photoionization dynamics \cite{ruberti_restricted_2019,ruberti2019onset,Ruberti2021,Ruberti2023}, the 3D time-dependent Schrödinger equation (TDSE) for the N-electron polyatomic para-aminophenol molecule interacting with the pump laser pulse 
\begin{equation}
i\hbar \frac{\partial |\Psi^{N}\left(t\right)\rangle}{\partial t} = \hat{H}^{N}_{Pump}\left(t\right) |\Psi^{N}\left(t\right)\rangle    
\label{adc_eq1}  
\end{equation}                           
is solved non-perturbatively, by expanding the TD many-electron wavefunction on the basis of the ground and excited RCS-ADC(n) states \cite{ruberti2019onset,Ruberti2021,schwickert_electronic_2022}
\begin{equation}
|\Psi^{N}\left(t\right)\rangle = \sum_{m,\mu}\,c_{m,\mu}\left(t\right) \hat{a}_{\mu}^{\dag} |\Phi^{N-1,RCS}_{m}\rangle + \sum_{I_{RCS}}\,c_{I_{RCS}}\left(t\right)|\tilde{\Psi}_{I_{RCS}}^{N}\rangle^{[n]} + c_{0}\left(t\right) |\Psi_{0}^{RCS}\rangle^{[n]}.   
\label{adc_eq2}       
\end{equation}

Here $|\Psi_{0}^{RCS}\rangle^{[n]}$ is the n-th order RCS correlated ground state, and $|\tilde{\Psi}_{I_{RCS}}^{N}\rangle^{[n]}$ indicate the excited intermediate states of the nth-order ADC(n) scheme for excitation of many-electron systems, built using the single-particle RCS basis. Within the first-order ADC(1) scheme ([n]=[1] in Eq.~\ref{adc_eq2}), the excited (N)-electron intermediate states span the configuration space consisting of one-hole – one-particle (1h1p)  ($|\tilde{\Psi}_{\alpha i}^{N}\rangle^{[n]}$) excitation classes built on top of the HF ground state \cite{Simpson2016,You2019}, while the extended second-order ADC(2)x scheme, which is the one used in this work, also includes the configuration space consisting of two-hole – two-particle (2h2p) ($|\tilde{\Psi}_{\alpha \beta i j}^{N}\rangle^{[n]}$) excitation classes built on top of the RCS correlated ground state \cite{ruberti_restricted_2019}. 

The first term on the right-hand side of Eq.~\ref{adc_eq2} describes the IS configuration states of B-spline RCS-ADC, which take the form of ionization channel-specific product states and reads $|\Psi^{N}_{\mu,m}\rangle = \hat{a}_{\mu}^{\dag} |\Phi^{N-1,RCS}_{m}\rangle$, where $\hat{a}_{\mu}^{\dag}$ is the creation operator of an electron in the IS molecular orbital $\psi_{\mu}$ and $|\Phi^{N-1,RCS}_{m}\rangle$ denotes the RCS-ADC ionic eigenstates. In the ADC(1) scheme, the ionic eigenstates consist of 1h states where one electron is removed from any of the 21 occupied HF molecular orbitals of neutral para-aminophenol. In the RCS-ADC(2)x scheme, the $|\Phi^{N-1,RCS}_{m}\rangle$ states are obtained by diagonalizing the ionic Hamiltonian calculated at the ADC(2)x level of theory and using the single-particle RCS basis set
\begin{equation}
\hat{H}_{ADC(2)x}^{(N-1,RCS)} |\Phi^{N-1,RCS}_{m}\rangle = I^{p}_{m} |\Phi^{N-1,RCS}_{m}\rangle\, ,
\label{adc_eq3}  
\end{equation}
where the ionization energy is given by $I^{p}_{m} = E_{m}^{(N-1)} - E_{0}^{RCS}$, $E_{0}^{RCS}$ is the RCS energy of the correlated ground state of the neutral molecule, and the ionic states are expanded into one-hole (1h), and two-holes - one-particle (2h1p) configurations derived from the RCS ground state 
\begin{equation}
|\Phi^{N-1,RCS}_{m}\rangle = \sum_{i} |\tilde{\Phi}_{i}^{N-1}\rangle + \sum_{\alpha i j} |\tilde{\Phi}_{\alpha i j}^{N-1}\rangle \,.             
\label{adc_eq4}  
\end{equation}
The ansatz of Eq.~\ref{adc_eq2} includes an explicit description of both electronic continua and many-body electron correlation effects, such as shakeup processes and breakdown of the molecular orbital (MO) picture. Moreover, the TD B-spline RCS-ADC theoretical framework fully accounts for the interchannel couplings between different ionization channels in the continuum, which can play an essential role during the ionization event. 

The total time-dependent Hamiltonian of Eq.~\ref{adc_eq1} reads 
\begin{equation}
\hat{H}^{N}_{Pump}\left(t\right) = \hat{\tilde{H}}^{N}_{RCS-ADC} + \hat{D}^{N}_{RCS-ADC} E_{Pump}\left(t\right) - i \hat{W}_{CAP}\,,                  
\label{adc_eq5}    
\end{equation}     
where a quadratic complex absorbing potential (CAP) $\hat{W}_{CAP}\left(r\right) = \eta \left(r-R_{CAP}\right)^{2}$ is used to eliminate wave packet reflections from the boundaries of the B-spline radial grid. The laser-molecule interaction is described within the dipole approximation in length form, and $\hat{\tilde{H}}^{N}_{RCS-ADC}$ and $\hat{D}^{N}_{RCS-ADC}$ are the representation of the shifted field-free Hamiltonian $\hat{\tilde{H}} = \hat{H} - E_{0}^{RCS}$ and the dipole operator $\hat{D}$, respectively, in the space of RCS-ADC intermediate states (see Eq. (2)). The values of the CAP parameters used are $\eta = 5 \times 10^{-3}$ and $R_{CAP} = 55$ a.u. 

The time propagation of the unknown coefficients of the many-electron wavefunction in Eq.~\ref{adc_eq2} is performed by means of the Arnoldi/Lanczos algorithm \cite{ruberti_full_2018,Ruberti2018HHG,ruberti2019onset}. During the simulation of the pump step, we have included in the expansion of the many-electron wavefunction only the ionization channels with an (ionization) spectral intensity value (see refs. \cite{ruberti_restricted_2019,ruberti2019onset,schwickert_electronic_2022}) greater than 1 $\%$. These states will be denoted in the following as $|\Phi^{N-1,RCS}_{m,Pump-ionized}\rangle$. The \textit{ab initio} simulation of the pump step has been performed using a linearly polarized pulse, characterized by a Gaussian temporal envelope, a central frequency $\hbar\omega_{Pump} = 263$ eV, peak intensity $I_{Pump} = 5 \times 10^{12}$ W/cm$^{2}$ and bandwidth $\simeq$ 3 eV (FWHM). Converged results have been obtained by using a 0.5 attosecond time-step and a 40-dimensional Krylov-space in the Arnoldi/Lanczos time propagation. 

The electron dynamics, both during and after the interaction of para-aminophenol with the pump pulse, is obtained from the time-dependent reduced ionic density matrix (R-IDM)  $\hat{\rho}^{R-IDM}\left(t\right)$ of the molecular ion, which is calculated by tracing out the unobserved photoelectron degree of freedom from the total time-dependent density matrix of the charge-neutral N-electron system 
\begin{equation}
\hat{\rho}^{R-IDM}\left(t\right) = Tr_{\mu}  \left[\langle\Psi^{N}\left(t\right) |\Psi^{N}\left(t\right)\rangle\right]\,.
\label{adc_eq6}    
\end{equation}                                           
Within the TD B-spline RCS-ADC(n) framework, and using Eq.~\ref{adc_eq2}, the resulting R-IDM on the basis of RCS-ADC(2)x ionic eigenstates reads 
\begin{eqnarray}
&\rho_{m,n}^{R-IDM}\left(t\right) = \sum_{\mu}  c_{m,\mu}\left(t\right) \left[
c_{m,\mu}\left(t\right) \right]^{*} +2e^{i\left(I_n^p-I_m^p\right)t}\times  \nonumber \\ 
& \int_{\infty}^{t} \sum_{\mu,\nu} w_{\nu,\mu} c_{m,\mu}\left(t^{'}\right) \left[c_{n,\nu}\left(t^{'}\right)\right]^{*} e^{-i\left(I_n^p-I_m^p\right)t^{'}} dt^{'},
\label{adc_eq7} 
\end{eqnarray}
where the latter term corrects for the loss of norm introduced by the CAP \cite{ruberti2019onset}, $I^{p}_{m}$ is the ionization potential of the ionic state m and $\omega_{\nu,\mu} $ is the CAP matrix element between photoelectron IS orbitals $\psi_{\nu}$ and $\psi_{\mu}$. From hereon in we shall omit the R-IDM superscript and denote the reduced ionic density matrix as $\rho_{m,n}$.
The time-dependent population of the ionic eigenstates is given by the diagonal elements 
\begin{equation}
P_{m}\left(t\right) = |\rho_{m,m}\left(t\right)|\,,      
\label{adc_eq8}  
\end{equation}
while the off-diagonal ones $\rho_{m,n}$ are related to the degrees of quantum electronic coherence \cite{ruberti_quantum_2022}, $0 \leq G_{m,n} \leq 1$, between any pair of populated ionic channels m and n 
\begin{equation}
 G_{m,n}\left(t\right) = \frac{|\rho_{m,n}\left(t\right)|}{\sqrt{P_{m}\left(t\right)\times P_{n}\left(t\right)}}\,.    
\label{adc_eq9} 
\end{equation}

\begin{figure}
\centering
\includegraphics[scale=1.2]{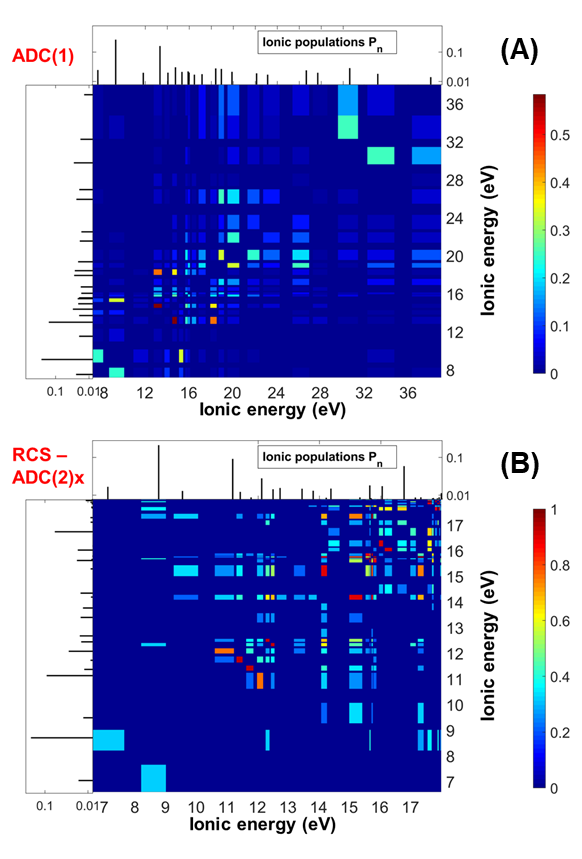}
\caption{\label{fig:adc_sm_1} \textit{Ab initio} TD B-spline RCS-ADC simulation of the pump-induced first ionization of para-aminophenol. The upper panel (A) shows the degrees of quantum electronic coherence $G_{m,n}$ (Eq. (9)) produced between each pair of ADC(1) cationic eigenstates populated by the pump pulse. The populations of the ionic eigenstates are also shown in the vertical and horizontal side panels. The lower panel (B) shows the degrees of quantum electronic coherence $G_{m,n}$ (Eq.~\ref{adc_eq9}) produced between each pair of bound RCS-ADC(2)x cationic eigenstates populated by the pump pulse. The populations of the ionic eigenstates, with energies $I^{p}_{m} < E_{DIP}$, are also shown in the vertical and horizontal side panels.}
\end{figure}

Fig.~\ref{fig:adc_sm_1} shows the calculated 2D map of quantum electronic coherences $ G_{m,n}$ between the ionic states of para-aminophenol, as it results from the TD B-spline RCS-ADC(1) (Fig.~\ref{fig:adc_sm_1} A) and RCS-ADC(2)x (Fig.~\ref{fig:adc_sm_1} B) simulations respectively. The final stationary populations $P_{m}\left(\infty \right)$ of the ADC(1) and the bound ADC(2)x-calculated eigenstates of the para-aminophenol cation are also shown in the vertical and horizontal side panels of Fig.~\ref{fig:adc_sm_1}. In the calculations presented here, all the ionic time-dependent populations $P_{m}\left(t \right)$ have converged to their final stationary value at $t \approx 100$ attoseconds after pump ionization. The ADC photoelectron spectrum shown in Fig.~\ref{fig:sim_res} A is obtained by using the t-surf technique \cite{Bray2021}, and using a t-surf radius $R_{t-surf} = 40$ a.u.

\subsubsection{Simulation of x-ray attosecond probe absorption}

We explicitly simulated the interaction of the pump-ionized system with the probe pulse and calculated the time-delay dependence of probe-induced x-ray absorption from the pump-ionized para-aminophenol cation in order to obtain a realistic description of the observable measured in the experiment. 

The formal validity of the presented theoretical results relies on non-overlapping pump and probe pulses. In practice, this limits the pump-probe time-delays that can be accurately described to the ones greater than half the duration of the x-ray pump pulse, i.e. in this case to delay times larger than $\approx 0.7$ femtoseconds.
In the ADC calculation, delays for which the pump and probe pulse are overlapped can be approximated by artificially truncating the pump pulse, meaning the pump and probe pulse are never simultaneously present in the calculation.
The non-perturbative, time-dependent description of the cation-probe interaction is performed by solving the time-dependent von Neumann equations \cite{Ruberti2021,schwickert_electronic_2022} for the characterized reduced ionic density matrix (Eq. (7)) interacting with the probe laser
\begin{equation}
\frac{d \hat{\rho} \left(t\right)}{dt} = \frac{i}{\hbar} [\hat{H}^{N-1}_{Probe}\left(t\right),\hat{\rho}\left(t\right)].
\label{adc_eq10} 
\end{equation}    

The probe-cation interaction is also described within the dipole approximation in length form. Eq.~\ref{adc_eq10} is solved in this work by representing the ionic density matrix both in the valence-ionized states populated by the pump process $|\Phi^{N-1,RCS}_{m,Pump-ionized}\rangle$, and the core-ionized states $|\Phi^{N-1,RCS}_{m^{'},Core}\rangle$, which are accessible to the probe pulse but were not populated in the pump ionization step. 
The probe pulse used in the simulation is characterized by a central photon energy $\hbar\omega_{Probe} = 526$ eV, peak intensity $I_{Pump} = 10^{12}$ W/cm$^{2}$ and bandwidth $\simeq$ 5 eV (FWHM).
The 526 eV photon energy can induce transitions from the valence-ionized states to core-ionized resonances $|\Phi^{N-1,RCS}_{m^{'},Core}\rangle$ of the para-aminophenol cation in the energy range of the oxygen $K$-edge, which are described here at the RCS-ADC(3) level of theory and employing the core-valence approximation. 
Finally, the measurable XAS signal resulting from the interaction of the x-ray probe pulse with the pump-prepared cationic system is obtained by the final population of the O(1s) core-ionized resonances $|\Phi^{N-1,RCS}_{m^{'},Core}\rangle$ upon interaction with the attosecond probe pulse. 

\begin{figure}
\centering
\includegraphics[scale=1.0]{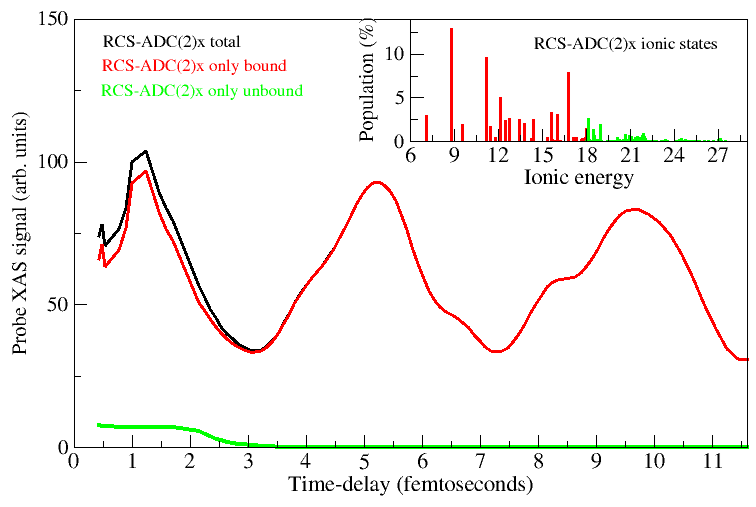}
\caption{\label{fig:adc_sm_3} The RCS-ADC calculated time-resolved XAS signal, integrated in the photon energy region from 521 to 530 eV. The red and green curve indicate the contribution of the bound (red sticks in the inset) and unbound (green sticks in the inset) RCS-ADC(2)x cationic states populated by the pump pulse, respectively. The total contribution of both bound and unbound states is shown by the black curve.}
\end{figure}

In order to model the ultrafast super-Coster-Kronig (CK) decay of the cationic states with an energy above the double ionization potential (DIP) energy threshold, in the probe simulations we introduced a phenomenological, exponentially-decaying term to the ionic Hamiltonian, namely $e^{-\frac{t}{\tau_{CK}}}$. The decay lifetimes used in the simulation, which have been calculated \textit{ab initio} by means of the Fano-ADC(2)x method \cite{kolorenc_fano-adc22_2020}, are: 
$\tau_{CK} = 345$ attoseconds for ionic states with energy above the DIP, but below 23 eV; $\tau_{CK} = 425$ attoseconds for states with energy between 23 eV and 30 eV; $\tau_{CK} = 864$ attoseconds for states with energy between 30 eV and 34 eV.
As a result, in the TD RCS-ADC simulation the population of the ionic states with energy above $E_{DIP}$ is driven by the aforementioned dumping factor and, at the same time, by the pump laser that tends to increase it for short time-delays below 1.6 femtoseconds. 
However, it is important to note that also for time-delays where there is considerable overlap between the pump and probe pulses, the present theoretical modeling only considers independent pump and probe effects, which could influence the accuracy of the theoretical points for such short time-delays. 

\begin{figure}
\centering
\includegraphics[scale=1.0]{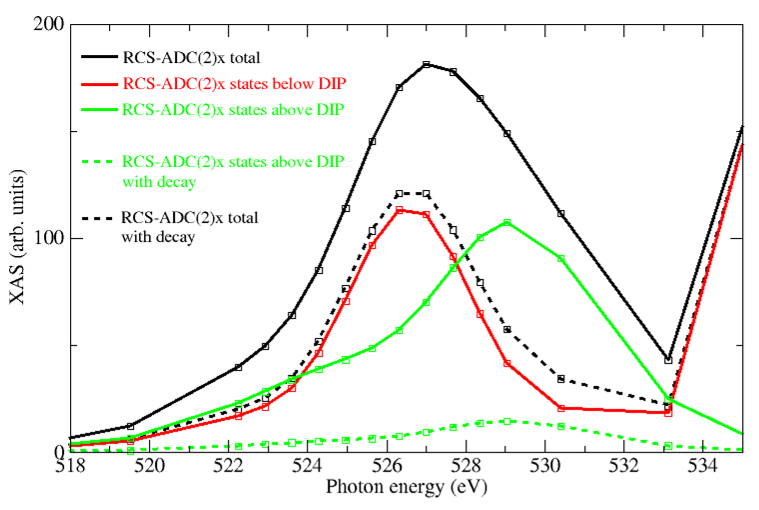}
\caption{\label{fig:adc_sm_2} The RCS-ADC calculated XAS signal in the photon energy region from 518 to 535 eV. The red and green curve indicate the contribution of the RCS-ADC(2)x cationic states populated by the pump pulse below and above the DIP, respectively. The total contribution of both set of states is shown by the black curve. Full lines do not include the autoionization decaying factor in the simulation, while dashed lines do.}
\end{figure}

Fig.~\ref{fig:adc_sm_2} shows the RCS-ADC calculated XAS signal in the photon energy region from 518 to 535 eV. Both contributions of the RCS-ADC(2)x cationic states populated by the pump pulse below and above the DIP is shown. 

Fig.~\ref{fig:adc_sm_3} shows the RCS-ADC calculated time-resolved XAS signal, integrated in the photon energy region from 521 to 530 eV. Both contributions of the RCS-ADC(2)x cationic states populated by the pump pulse below and above the DIP is also shown. 

The contribution to the XAS signal from the decayed cationic states is converted into absorption from the resulting dicationic populations (which will partially contribute to the same photon energy absorption window). The total trXAS signal can therefore be therefore decomposed into two contributions:
\begin{equation}
XAS^{TOT}\left(t\right) = XAS^{Cation}\left(t\right) + 
\end{equation}
where $XAS^{Cation}\left(t\right)$ describes the contribution from the cationic states below DIP and above DIP,
\begin{equation}
XAS^{Cation}\left(t\right) = XAS^{Cation}_{Below DIP}\left(t\right) + XAS^{Cation}_{Above DIP}\left(t\right)\,,
\end{equation}
and 
\begin{equation}
XAS^{Dication}\left(t\right) = \alpha \times XAS^{Decayed Cation}\left(t\right) 
\end{equation}
describes the dication contribution to the XAS signal at time t as a percentage of the absorption contribution of the cationic states that have already undergone super-CK decay at time t. In other words, the contribution of the unbound ionic states is added back, as it decays (see the green curve in Fig.~\ref{fig:adc_sm_2}), multiplied by a factor $\alpha$ to the total XAS signal. The case $\alpha = 1$ corresponds to a scenario in which the overall dicationic population absorbs into the probe photon energy window (523-531 eV in the ADC calculation) exactly as much as the cationic states that have decayed (super-Coster-Kronig) would have done. 

In Fig.~\ref{fig:sim_res}, an additional exponential decay factor is added to the calculated XAS signal in order to model a shift of the absorption window driven by nuclear motion, which should indeed lead to a decay of the background XAS signal.

\end{document}

%% file: authors.tex
\author[1,2]{{Taran} {Driver}\thanks{Authors Contributed Equally}\thanks{tdriver@stanford.edu}}
\author[3,4]{{Zhaoheng} {Guo}$^*$}
\author[1,2,3]{{Erik} {Isele}$^*$}
\author[5,6]{{Gilbert} {Grell}$^*$}
\author[7]{{Marco} {Ruberti}$^*$}
\author[1,2,8]{{Jordan T.} {O'Neal}}

\author[7]{{Oliver} {Alexander}}
\author[9,1]{{Sandra} {Beauvarlet}}
\author[4]{{David} {Cesar}}
\author[4]{{Joseph} {Duris}}
\author[1,2]{{Douglas} {Garratt}}
\author[1,2]{{Kirk A.} {Larsen}}
\author[4]{{Siqi} {Li}}
\author[10]{{P\v{r}emysl} {Koloren\v{c}}}
\author[11]{{Gregory A.} {McCracken}}
\author[11]{{Daniel} {Tuthill}}
\author[11]{{Zifan} {Wang}}

\author[9]{{Nora} {Berrah}}
\author[12,13]{{Christoph} {Bostedt}}
\author[2,14]{{Kurtis} {Borne}}
\author[2]{{Xinxin} {Cheng}}
\author[11]{{Louis F.} {DiMauro}}
\author[15]{{Gilles} {Doumy}}
\author[3,4]{{Paris L.} {Franz}}
\author[2]{{Andrei} {Kamalov}}
\author[2]{{Xiang} {Li}}
\author[2]{{Ming-Fu} {Lin}}
\author[2]{{Razib} {Obaid}}
\author[6]{{Antonio} {P\'icon}}
\author[3,4]{{River R.} {Robles}}
\author[14]{{Daniel} {Rolles}}
\author[14]{{Artem} {Rudenko}}
\author[16]{{Moniruzzaman} {Shaikh}}
\author[16]{{Daniel S.} {Slaughter}}
\author[4]{{Nicholas S.} {Sudar}}
\author[1,2,3]{{Emily} {Thierstein}}
\author[17]{{Kiyoshi} {Ueda}}
\author[14]{{Enliang} {Wang}}
\author[1,2,3]{{Anna L.} {Wang}}
\author[16]{{Thorsten} {Weber}}
\author[1,2]{{Thomas J. A.} {Wolf}}
\author[15,18]{{Linda} {Young}}
\author[4]{{Zhen} {Zhang}}

\author[7]{{Vitali} {Averbukh}} 
\author[16]{{Oliver} {Gessner}}
\author[1,3,8]{{Philip H.} {Bucksbaum}}
\author[2,19]{{Matthias F.} {Kling}}
\author[6]{{Alicia} {Palacios}}
\author[6]{{Fernando} {Mart\'in}\thanks{fernando.martin@uam.es}}
\author[7]{{Jon P.} {Marangos}\thanks{j.marangos@imperial.ac.uk}}

\author[2]{{Peter} {Walter}}

\author[2,4,19]{{Agostino} {Marinelli}\thanks{marinelli@slac.stanford.edu}}
\author[1,2]{{James P.} {Cryan}\thanks{jcryan@slac.stanford.edu}}

\affil[1]{{Stanford PULSE Institute},
    {SLAC National Accelerator Laboratory},
    {{Menlo Park}, {CA}, {USA}}}

\affil[2]{{Linac Coherent Light Source},
    {SLAC National Accelerator Laboratory},
    {{Menlo Park}, {CA}, {USA}}}

\affil[3]{{Department of Applied Physics},
    {Stanford University},
    {{Stanford}, {CA}, {USA}}}

\affil[4]{{SLAC National Accelerator Laboratory},
    {{Menlo Park}, {CA}, {USA}}}

\affil[5]{{Instituto Madrile\~no de Estudios Avanzados en Nanociencia (IMDEA-Nanociencia)},
    {{Madrid}, {Spain}}}

\affil[6]{{Departamento de Qu\'imica},
    {Universidad Aut\'onoma de Madrid},
    {{Madrid}, {Spain}}}

\affil[7]{{The Blackett Laboratory}, 
    {Imperial College London}, 
    {{London}, {United Kingdom}}}

\affil[8]{{Department of Physics},
    {Stanford University},
    {{Stanford}, {CA}, {USA}}}

 \affil[9]{{Department of Physics}, 
    {University of Connecticut}, 
    {{Storrs}, {Connecticut}, {USA}}}

 \affil[10]{{Faculty of Mathematics and Physics, Institute of Theoretical Physics}, 
    {Charles University}, 
    {{Prague}, {Czechia}}}

\affil[11]{{Department of Physics}, 
    {The Ohio State University}, 
    {{Columbus}, {Ohio}, {USA}}}

\affil[12]{{Paul Scherrer Institute}, 
    {{Villigen}, {Switzerland}}}
    
\affil[13]{{LUXS Laboratory for Ultrafast X-ray Sciences, Institute of Chemical Sciences and Engineering}, 
    {Ecole Polytechnique Federale de Lausanne (EPFL)},
    {{Lausanne}, {Switzerland}}}

\affil[14]{{J.R. Macdonald Laboratory},
    {Kansas State University},
    {{Manhattan}, {Kansas}, {USA}}}

\affil[15]{{Chemical Sciences and Engineering Division},
    {Argonne National Laboratory},
    {{Lemont}, {IL}, {USA}}}

\affil[16]{{Chemical Sciences Division}, 
    {Lawerence Berkeley National Laboratory}, 
    {{Berkeley}, {California}, {USA}}}

\affil[17]{{Department of Chemistry},
    {Tohoku University},
    {{Sendai}, {Japan}}}

\affil[18]{{Department of Physics and James Franck Institute},
    {The University of Chicago},
    {{Chicago}, {IL}, {USA}}}

\affil[19]{{Department of Photon Science},
    {SLAC National Accelerator Laboratory},
    {{Menlo Park}, {CA}, {USA}}}